\documentclass[fleqn,usenatbib]{mnras}

\usepackage[T1]{fontenc}

\DeclareRobustCommand{\VAN}[3]{#2}
\let\VANthebibliography\thebibliography
\def\thebibliography{\DeclareRobustCommand{\VAN}[3]{##3}\VANthebibliography}

\defcitealias{Murugeshan2020}{M20}
\defcitealias{ManceraPina2021}{MP21}

\usepackage{graphicx}	
\usepackage{amsmath}	
\usepackage{amssymb}	
\usepackage{rotating} 
\usepackage{pdflscape}

\usepackage[varvw]{newtxmath}
\usepackage{amssymb}
\usepackage{lipsum}
\usepackage{afterpage}
\usepackage{threeparttable}
\usepackage{placeins}
\usepackage{multirow,makecell,booktabs}
\usepackage{graphicx}
\usepackage{subfig}
\usepackage[nameinlink]{cleveref}

\Crefname{figure}{Fig.}{Figs.}
\Crefname{equation}{Eq.}{Eqs.}
\Crefname{section}{Sect.}{Sects.}
\usepackage{enumitem}

\usepackage{newtxtext,newtxmath}

\usepackage{tikz}
\usetikzlibrary{positioning, arrows.meta}

\newcommand{\hi}{\mbox{H{\sc i}}}

\newcommand{\Mbar}{$M_{\rm bar}$}
\newcommand{\jbar}{$j_{\rm bar}$}
\newcommand{\jgas}{$j_{\rm gas}$}
\newcommand{\jstar}{$j_{\star}$}

\newcommand{\dg}{^{\circ}}

\newcommand{\kms}{km\,s$^{-1}$}

\definecolor{darkgray}{rgb}{0.52, 0.52, 0.51}

\newcommand{\um}{\rm \mu m}

\newcommand{\HI}{\textsc{Hi}}

\newcommand{\wise}{{\it WISE}}

\newcommand{\sofia}{\textsc{SoFiA}}

\newcommand{\barolo}{\textsc{3d Barolo}}

\def\alphaj{0.54\pm0.08}
\def\cj{2.96\pm0.06}
\def\sigmaj{0.17\pm0.03}
\def\nuj{5.8\pm2.8}

\def\sizeman{151}
\def\alphaman{0.58\pm0.02}
\def\cman{2.75\pm0.02}
\def\sigmaman{0.20\pm0.01}

\def\sizemanlow{81}
\def\alphamanlow{0.69 \pm 0.04}
\def\cmanlow{2.90 \pm 0.06}
\def\sigmamanlow{0.21 \pm 0.02}

\def\sizemanhigh{70}
\def\alphamanhigh{0.50 \pm 0.05}
\def\cmanhigh{2.78 \pm 0.03}
\def\sigmamanhigh{0.19 \pm 0.02}

\def\sizemur{109}
\def\alphamur{0.53\pm0.02}
\def\cmur{2.84\pm0.02}
\def\sigmamur{0.18\pm0.01}

\def\sizemurlow{44}
\def\alphamurlow{0.67 \pm 0.07}

\def\sizemurhigh{65}
\def\alphamurhigh{0.54 \pm 0.06}
\def\cmurhigh{2.83 \pm 0.03}
\def\sigmamurhigh{0.19 \pm 0.02}

\def\sizekur{11}
\def\alphakur{0.83 \pm 0.08}
\def\ckur{3.28 \pm 0.12}
\def\sigmakur{0.15 \pm 0.04}

\def\sizebut{13*}
\def\alphabut{0.68 \pm 0.09}
\def\cbut{3.10 \pm 0.14}
\def\sigmabut{0.20 \pm 0.05}

\def\sizeuma{15**}
\def\alphauma{0.32 \pm 0.09}
\def\cuma{2.83 \pm 0.05}
\def\sigmauma{0.17 \pm 0.04}

\title[Angular momentum of AMIGA galaxies]{The AMIGA sample of isolated galaxies -- Effects of Environment on Angular momentum}

\author[A Sorgho et al.]{
A. Sorgho$^{1}$,\thanks{E-mail: asorgho@iaa.es}
L. Verdes-Montenegro$^{1}$,
K. M. Hess$^{1,2}$,
M. G. Jones$^{3}$,
T. H. Jarrett$^{4}$,
\newauthor
S. Sanchez-Exp\'osito$^{1}$,
J. Garrido$^{1}$
\\
$^{1}$Instituto de Astrof\'isica de Andaluc\'ia (IAA-CSIC), Glorieta de la Astronom\'ia s/n, 18008 Granada, Spain\\
$^{2}$ Department of Space, Earth and Environment, Chalmers University of Technology, Onsala Space Observatory, SE-43992 Onsala, Sweden\\
$^{3}$Steward Observatory, University of Arizona, 933 North Cherry Avenue, Rm. N204, Tucson, AZ 85721-0065, USA\\
$^{4}$Astronomy Department, University of Cape Town, Private Bag X3, Rondebosch 7701, South Africa\\
}

\date{Accepted 2023 December 20}

\begin{document}
\label{firstpage}
\pagerange{\pageref{firstpage}--\pageref{lastpage}}
\maketitle

\begin{abstract}
We investigate the relationship between the baryonic angular momentum and mass for a sample of 36 isolated disc galaxies with resolved \hi\ kinematics and infrared {\it WISE} photometry drawn from -- and representative in terms of morphologies, stellar masses and \hi-to-star fraction of -- the carefully-constructed AMIGA sample of isolated galaxies. Similarly to previous studies performed on non-isolated galaxies, we find that the relation is well described by a power law $j_{\rm bar} \propto M_{\rm bar}^\alpha$. We also find a slope of $\alpha = \alphaj$ for the AMIGA galaxies, in line with previous studies in the literature; however, we find that the specific angular momenta of the AMIGA galaxies are on average higher than those of non-isolated galaxies in the literature. This is consistent with theories stipulating that environmental processes involving galaxy-galaxy interaction are able to impact the angular momentum content of galaxies. However, no correlation was found between the angular momentum and the degree of isolation, suggesting that there may exist a threshold local number density beyond which the effects of the environment on the angular momentum become important.
\end{abstract}




\section{Introduction}
Viewed as a basic property of galaxies, the angular momentum holds an important place in constraining theories of galaxy formation and evolution \citep{Fall1983,Fall2013}. Initial analytical studies on the subject proposed that angular momentum is acquired by the dark matter halo through tidal torques, during the proto-galactic formation phase \citep[see e.g.,][]{Peebles1969,Fall1980,White1984}. Additionally, since the baryonic matter in galaxies is thought to experience the same torque, its angular momentum is expected to follow the same distribution as the dark matter (DM) halo \citep[e.g.,][]{Mo1998}.

On the other hand, one of the most important aspects of the angular momentum in the context of the galaxy evolution study lies in its relationship with the mass. In the framework of the cold dark matter (CDM) cosmology, the angular momentum of the DM halo (characterised by the global spin parameter) is predicted to approximately be independent of the mass \citep[e.g.,][]{Barnes1987}, leading to a power-law relation between the DM's specific angular momentum $j_{\rm DM}$ (i.e, the angular momentum per unit mass) and its mass $M_{\rm DM}$: $j_{\rm DM} \propto M^\alpha_{\rm DM}$, with $\alpha\sim2/3$. This relation also holds for the baryons within the DM halo, since they are expected to follow the DM in the angular momentum distribution.

The total budget of a galaxy's baryonic angular momentum is essentially provided by the stellar and gas components making up the galaxy. The initial observational study of the $j{-}M$ relation on the stellar component \citep{Fall1983} found a slope similar to the theoretical prediction, but also revealed that at given stellar mass, disc galaxies have higher specific angular momenta than early type galaxies. Subsequent and more comprehensive studies refined these results, demonstrating the dependency of the angular momentum on galaxy morphological type \citep{Romanowsky2012,Fall2013}. More recently, several studies have included the gas component in the evaluation of angular momentum, providing a more complete estimate of the total baryonic content \citep[e.g,][]{Obreschkow2014,Obreschkow2016,Elson2017,Hardwick2022a,Romeo2023}. Although the emerging relation of the total baryonic angular momentum does not largely differ from that of the stellar component, the emerging picture suggests that complex mechanisms are responsible for the observed angular momentum content of galaxies. For example, the retained angular momentum fraction (i.e, the ratio between the baryonic and DM angular momenta) is presumably higher for galaxies with higher baryon fraction, suggesting that these galaxies conserve better their angular momentum during their formation phase \citep[e.g.,][]{Posti2018a,Romeo2023}.

Numerous theoretical studies have also attempted, over the recent years, to provide a complete description of how the angular momentum of the baryonic component varies over a galaxy's lifetime. Today, the generally accepted picture is that both internal and external processes (such as star formation, stellar feedback, gas inflow and outflow, merging) are capable of affecting the angular momentum of galaxies \citep[e.g.,][]{Danovich2015,Jiang2019}. This in turn can alter the position of individual galaxies in the $j{-}M$ plane.

While the occurrence and importance of internal mechanisms are independent of the environment, the external processes are significantly impacted by local density in the medium around galaxies. In fact, several studies on galaxy formation and evolution have shown that environment plays an important role in shaping the physical properties of galaxies \citep[e.g.,][]{Dressler1980,Haynes1984,Cayatte1990,Goto2003}. From a morphological point of view, the neutral hydrogen (\hi) content is arguably among the most important parameters in tracing environmental processes, since it constitutes the envelope that is most affected by said processes \citep[see e.g,][]{Chung2007} and the reservoir of gas out of which stars are formed (via molecular gas). Galaxies evolving in dense environments tend to be more \hi\ deficient than their counterparts in low-density regions \citep[e.g.,][]{Giovanelli1985,Solanes2001,Verdes-Montenegro2001,Boselli2006}. On the other hand, galaxies residing in the lowest density environments are less exposed environmental processes: their \hi\ content is higher than the average, while their \hi\ distribution is more orderly \citep{Espada2011,Jones2018}.

Most observational investigations since the original \citet{Fall1983} study have focused on either providing a better constraint of the $j{-}M$ relation with respect to morphological type and gas fraction, or reconciling measured the retained fraction of angular momentum with the numerical predictions \citep[see above references, but also][hereafter MP21]{Posti2018,ManceraPina2021}. However, little attention was given to the environmental dependency of the angular momentum distribution \citep[the few available studies include][hereafter M20]{Murugeshan2020}; in particular, no existing study provides analysis on galaxies selected in extremely low density environments.

In this work, we investigate the specific angular momentum of a subset of the AMIGA (Analysis of the interstellar Medium in Isolated GAlaxies) sample \citep{Verdes-Montenegro2005a}, the most carefully constructed sample of isolated galaxies available to date. The degree of isolation of galaxies in the catalogue was evaluated based on two main criteria: the local environment number density $\eta_k$ and the total force $Q$ exerted on the galaxies by their neighbours \citep{Verley2007b,Argudo-Fernandez2013}. More isolated than most of their field counterparts, the galaxies in AMIGA were found to be almost ``nurture free", exhibiting extremely low values for parameters that are usually enhanced by interaction \citep[e.g.,][]{Lisenfeld2007,Lisenfeld2011,Espada2011,Sabater2012}. Therefore, the sample provides, by definition, a good reference for evaluating the \jbar$-$\Mbar\ relation (the angular momentum - mass relation for the baryonic component) in interaction-free galaxies in the local Universe. The aim of the present investigation is to evaluate how the environment impacts the angular momentum of disc galaxies. Indeed, how environmental processes affect the angular momentum content of a galaxy is not straightforwards, with the change in $j$ being dependent on the specifications of the interactions. However, current simulations tend to agree that processes such as mergers could potentially redistribute the stellar angular momentum from the inner regions of galaxies out to their outer parts \citep[e.g.,][]{Navarro1994,Hernquist1995,Zavala2008,Lagos2017,Lagos2018}. It is therefore possible that galaxy interactions transfer part of the (stellar and gas) disc angular momentum into the DM halo, effectively reducing the ``observable'' angular momentum content. However, no observational study, to date, has conclusively shown evidence of this effect. If these theoretical predictions are correct, we then expect isolated galaxies to have retained a larger fraction of their initial angular momentum -- resulting in these galaxies having higher $j$ values. We therefore make use of the AMIGA sample to investigate this hypothesis, which is undoubtedly the best existing sample candidate for the study.

The paper is organised as follows: in \Cref{sec:data} we describe the AMIGA sample, the \hi\ and mid-infrared data used in the analysis. Next, we present details on the measurement of the specific angular momentum in \Cref{sec:j}. The relation between $j$ and the mass is then presented and analysed in \Cref{sec:jvsm}, with a discussion within the context of galaxy evolution in \Cref{sec:discussion}. Finally, we summarise and layout the future prospects in \Cref{sec:summary}.

\section{Data}\label{sec:data}

\subsection{The AMIGA Sample of Isolated galaxies}\label{sec:data:amiga}
The AMIGA \citep{Verdes-Montenegro2005a} galaxies were selected from the 1050 isolated galaxies of the CIG \citep[Catalogue of Isolated Galaxies,][]{Karachentseva1973} catalogue. The original study of \citet{Verdes-Montenegro2005a} found that the AMIGA sample has properties as close as possible to field galaxies, with an optical luminosity function representative of the lower density parts of galaxy environments. The study also performed a completeness test and concluded that the sample was over 80\% complete for objects with B-band magnitudes brighter than 15.0 and within 100 Mpc. The morphological study of the sample revealed that it contains 14\% of early-type (E/S0) galaxies, with a vast majority of the galaxies (82\%) ranging from Sa to Sd Hubble types \citep{Sulentic2006}. Several multi-wavelength studies have since then refined the AMIGA sample to ensure that it is as ``nurture-free'' as possible, by eliminating galaxies that are suspected to have undergone recent interaction. In particular, \citet{Verley2007a} mapped the projected neighbours of 950 CIG galaxies with systemic velocities higher than 1500 \kms, down to a B magnitude limit of 17.5, and within within a radius of 0.5 Mpc around each of these galaxies. The velocity cut ensures that nearby galaxies -- i.e, those closer than 20 Mpc -- are not included in the AMIGA sample since their low distance would result in impractically large searching areas for potential neighbours during the evaluation of the isolation degree.
In their study, the authors identified only 636 galaxies that appeared to be isolated. Subsequently, \citep{Verley2007b} estimated the influence of their potential neighbours on the CIG galaxies by measuring their local number density $\eta_k$\footnote{by definition, $\eta_k$ can only be determined for galaxies having at least two neighbours.} and the tidal strength $Q$ to which they are subject, providing a tool for quantifying the degree of isolation of the sample galaxies. These isolation parameters allowed the authors to i) find that the 950 galaxies of $v>1500$ \kms\ presented a continuous spectrum of isolation, ranging from strictly isolated to mildly interacting galaxies, and to ii) produce a subsample of the 791 most isolated AMIGA galaxies. These isolated galaxies were selected such that $\eta_k < 2.4$ and $Q < -2$. Although the isolation criteria were later revised by \citet{Argudo-Fernandez2013} who further reduced the sample size to 426 galaxies\footnote{the authors end up with a smaller sample because not all CIG galaxies are in the SDSS footprint.} based on photometric and spectroscopic data from the SDSS Data Release 9, there is agreement that the \citet{Verley2007b}'s sample of 791 galaxies provides a suitable nurture-free baseline for effectively quantifying the effects of galaxy interactions \citep[][also see discussion in \Cref{sec:jvsm:isol}]{Leon2008,Sabater2008,Lisenfeld2011,Jones2018}: we will hereafter refer to this sample as the Verley07b sample.

\begin{figure*}
    \centering
    \includegraphics[width=\textwidth]{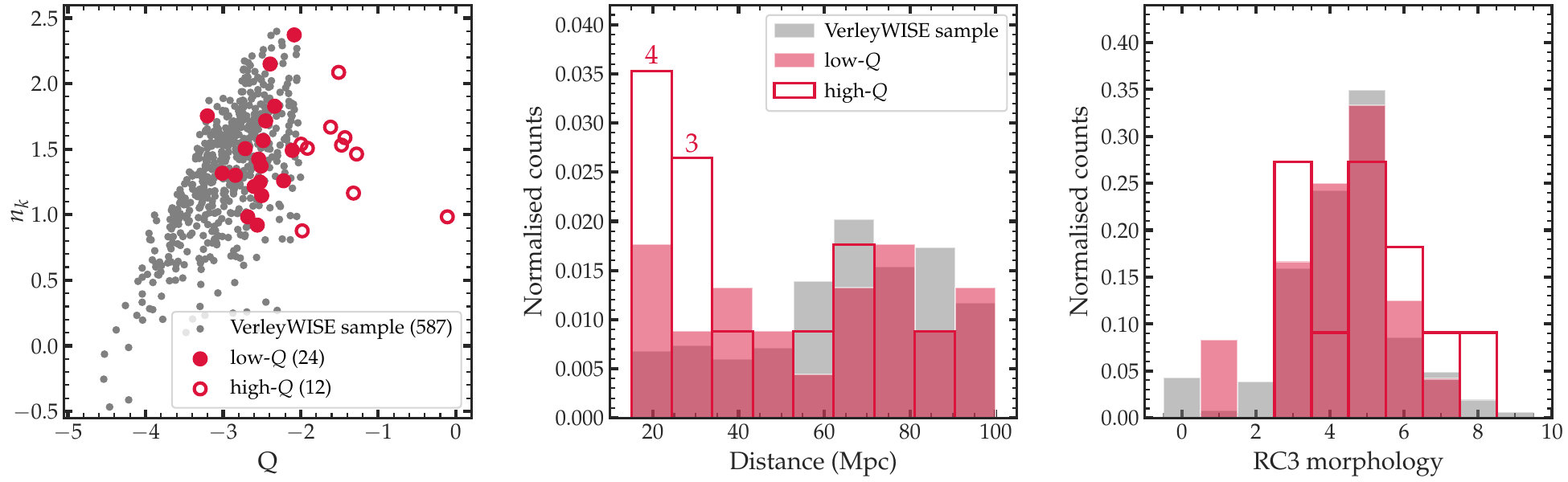}
    \vspace{-20pt}
    \caption{Comparison between the {\it VerleyWISE} and angular momentum samples. {\it left}: the local number density as a function of the tidal forces parameter; the bracketed numbers in the legend indicate the sample sizes. {\it Middle}: distribution of the heliocentric distances; {\it right}: distribution of the RC3 morphologies. The histograms of the middle and right panels were normalised {\it per sample}, and are therefore not indicative of the relative sizes of the individual samples. For reference, the numbers above the first two bars of the middle panel represent the counts of the low-$Q$ sample in the corresponding bins.}\label{fig:amiga-sub}
\end{figure*}

From the initial sample of 950 galaxies, we selected 38 galaxies for which high-quality \hi\ data are available (see \Cref{sec:data:hi}). Among these, 36 galaxies (except CIG 587 \& 812) were further detected in mid-infrared (see \Cref{sec:data:wise}): only these galaxies will be considered in the angular momentum analysis below, and will be referred to as the angular momentum sample (or $j$-sample). From this sample, 24 meet the isolation criteria of \citet{Verley2007b}, while the remaining 12 were classified by the authors as non-isolated. A closer look at the distribution of the $j$-sample's isolation parameters reveals that the groups of 24 and 12 galaxies are rather separated by the tidal force $Q$ (left panel of \Cref{fig:amiga-sub}): we will therefore refer to them as the low-$Q$ and high-$Q$ samples respectively in the next sections.

\begin{figure}
\centering
\begin{tikzpicture}[
squarednode/.style={rectangle, rounded corners=3pt, draw=black, fill=gray!5, minimum size=10mm, align=center},
textnode/.style={rectangle, rounded corners=2pt, draw=gray, midway, dash pattern=, fill=white, minimum size=5mm, align=center}
]
\node[squarednode]  (cig)    {{\bf CIG sample} \\ 1050 galaxies};
\node[squarednode]  (highv)       [below=1.5cm of cig] {{\bf high $v$ sample} \\ 950 galaxies};
\node[squarednode]  (v07a)       [right=of highv] {{\bf Verley+ 07a} \\ 636 galaxies};
\node[squarednode]  (arg13)       [left=of highv] {{\bf Argudo-}\\ {\bf Fern\'andez+ 13} \\ 426 galaxies};
\node[squarednode]  (v07b)       [below left=2cm and 0.5cm of highv] {{\bf Verley+ 07b} \\ 791 galaxies \\ (most isolated)};
\node[squarednode]  (highq)       [below right=2cm and 0.5cm of highv] {{\bf high-$Q$ sample} \\ 12 galaxies};
\node[squarednode]  (opt)       [below=2cm of v07b] {{\bf VerleyWISE sample} \\ 587 galaxies};
\node[squarednode]  (js)       [below left=0.5cm and 0.7cm of highq] {{\bf $j$-sample} \\ 36 galaxies};
\node[squarednode]  (lowq)  [right=4cm of opt] {{\bf low-$Q$ sample} \\ 24 galaxies};

\draw[-{Stealth[length=3mm, width=2mm]}] (cig.south) -- (highv.north) node[textnode] {$v \geq 1500$ \kms};
\draw[-{Stealth[length=3mm, width=2mm]}] (highv.east) -- (v07a.west);
\draw[-{Stealth[length=3mm, width=2mm]}] (highv.west) -- (arg13.east);
\draw[-{Stealth[length=3mm, width=2mm]}] ([xshift=-.5cm]highv.south) -- (v07b.north) node[textnode] {$\eta_k < 2.4$ \\ $Q < -2$};
\draw[-{Stealth[length=3mm, width=2mm]}] (v07b.south) -- (opt.north) node[textnode] {\wise\ photometry \\ + \hi\ mass};
\draw[-{Stealth[length=3mm, width=2mm]}] ([xshift=.5cm]highv.south) -- (highq.north) node[textnode] {\wise\ photometry\\ + \hi\ mass\\ + \hi\ kinematics};
\draw[-{Stealth[length=3mm, width=2mm]}] (opt.east) -- (lowq.west) node[textnode] {\hi\ kinematics};
\draw[-{Stealth[length=3mm, width=2mm]}] ([xshift=-.5cm]lowq.north) -- ([yshift=-.1cm]js.east);
\draw[-{Stealth[length=3mm, width=2mm]}] (highq.south) -- ([yshift=.1cm]js.east);
\end{tikzpicture}


\vspace{-20pt}
\caption{The samples selection process, from the overall CIG sample to the angular momentum sample (or $j$-sample).}\label{fig:diagram}
\end{figure}
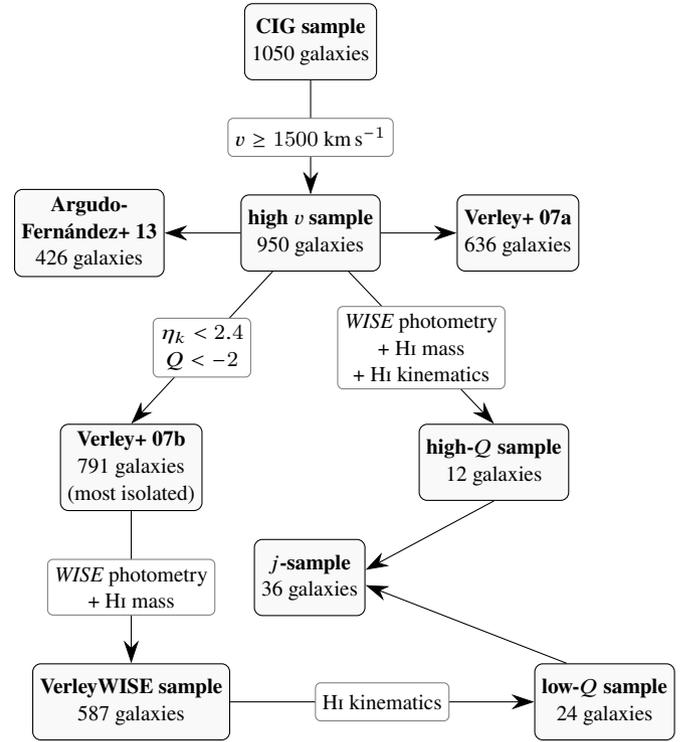

To assess how representative the $j$-sample is of the larger AMIGA sample of isolated galaxies, we further constrain the AMIGA sample to those galaxies for which we can reliably determine both the stellar and \hi\ properties. Among the 791 galaxies in the Verley07b sample, only 587 galaxies have both their \hi\ masses and \wise\ \citep[Wide-field Infrared Survey Explorer;][]{Wright2010} infrared photometry available (see \Cref{sec:data:wise}). We refer to these 587 galaxies as the {\it VerleyWISE} sample. The different samples are summarised in the diagram of \Cref{fig:diagram}. In \Cref{fig:amiga-sub} we compare the distribution of the isolation parameters, distance and morphologies in both the angular momentum and VerleyWISE samples. In terms of isolation, the galaxies in the low-$Q$ sample occupy the same parameter space as the VerleyWISE sample although their values of the $Q$ parameter tend to be on the upper end of the VerleyWISE sample. Furthermore, the distances and morphologies of the low-$Q$ sample appear to be distributed similarly to those of the VerleyWISE sample. On the other hand, while the high-$Q$ sample's morphologies are distributed roughly similar to those of the VerleyWISE sample, its distance distribution is skewed towards the lower limit: 7 out of the 12 galaxies in the sample are closer than 40 Mpc, while the median distances of the other two samples are in the range $\sim60-80$ Mpc.

\begin{figure}
    \centering
    \includegraphics[width=\columnwidth]{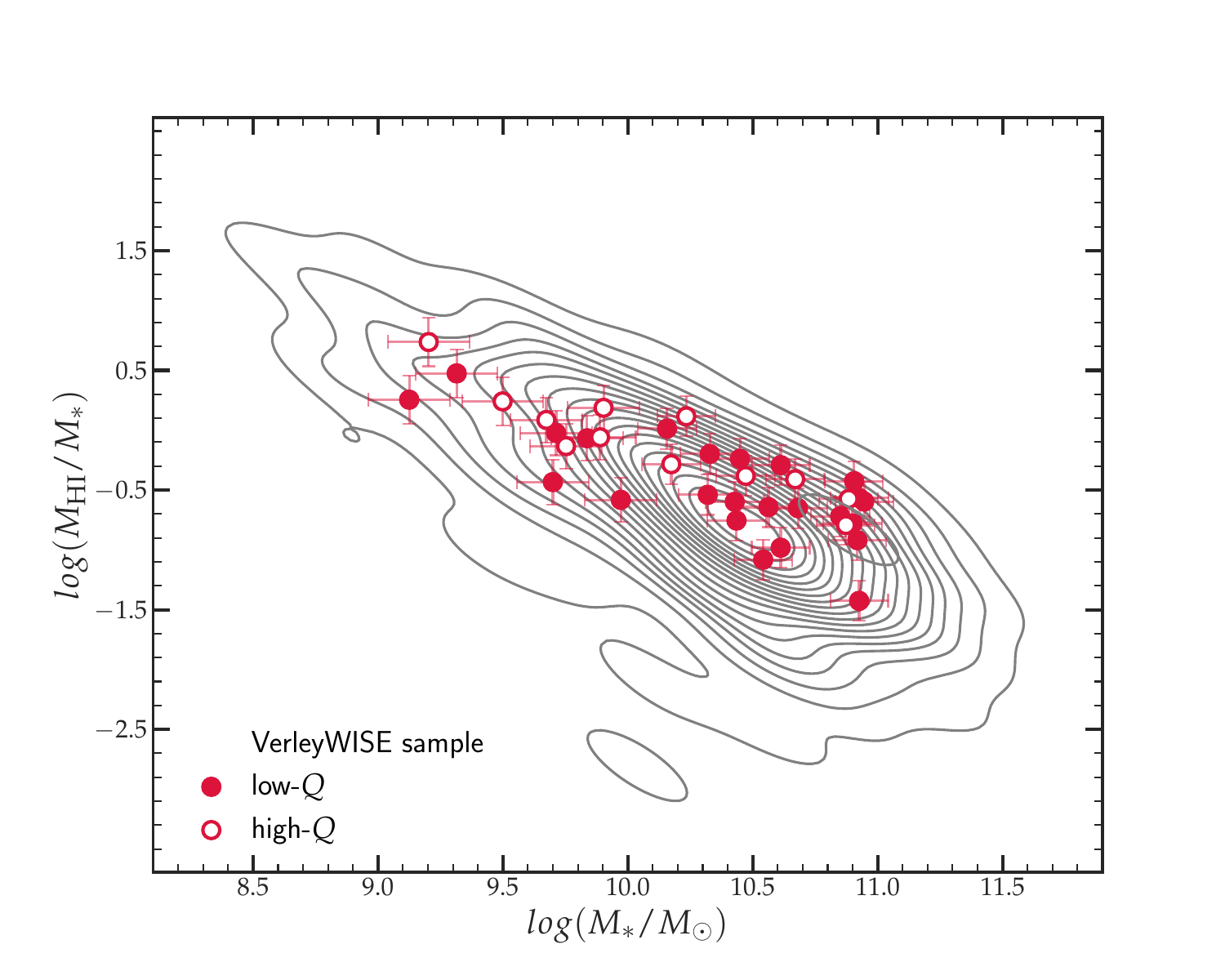}
    \vspace{-20pt}
    \caption{\hi-to-stellar mass fraction as a function of the stellar mass for the AMIGA VerleyWISE ({\it contours}) and sub- ({\it stars}) samples.}\label{fig:mvsmfrac}
\end{figure}

As for the trends of the stellar and \hi\ masses, \Cref{fig:mvsmfrac} shows that the isolated $j$-sample follows the distribution of the VerleyWISE sample. In fact, the \hi-to-stellar mass fractions of the galaxies in both the low-$Q$ and high-$Q$ samples are distributed uniformly across the stellar mass range, residing together with the majority of the VerleyWISE sample galaxies in the parameter space. Moreover, unlike the distance parameter, the \hi\ mass fractions of the high-$Q$ sample present no discrepancy with those of the low-$Q$, although the high-$Q$ galaxies tend to have higher \hi\ mass fractions than those in the low-$Q$ sample. 

\begin{figure}
    \centering
    \includegraphics[width=\columnwidth]{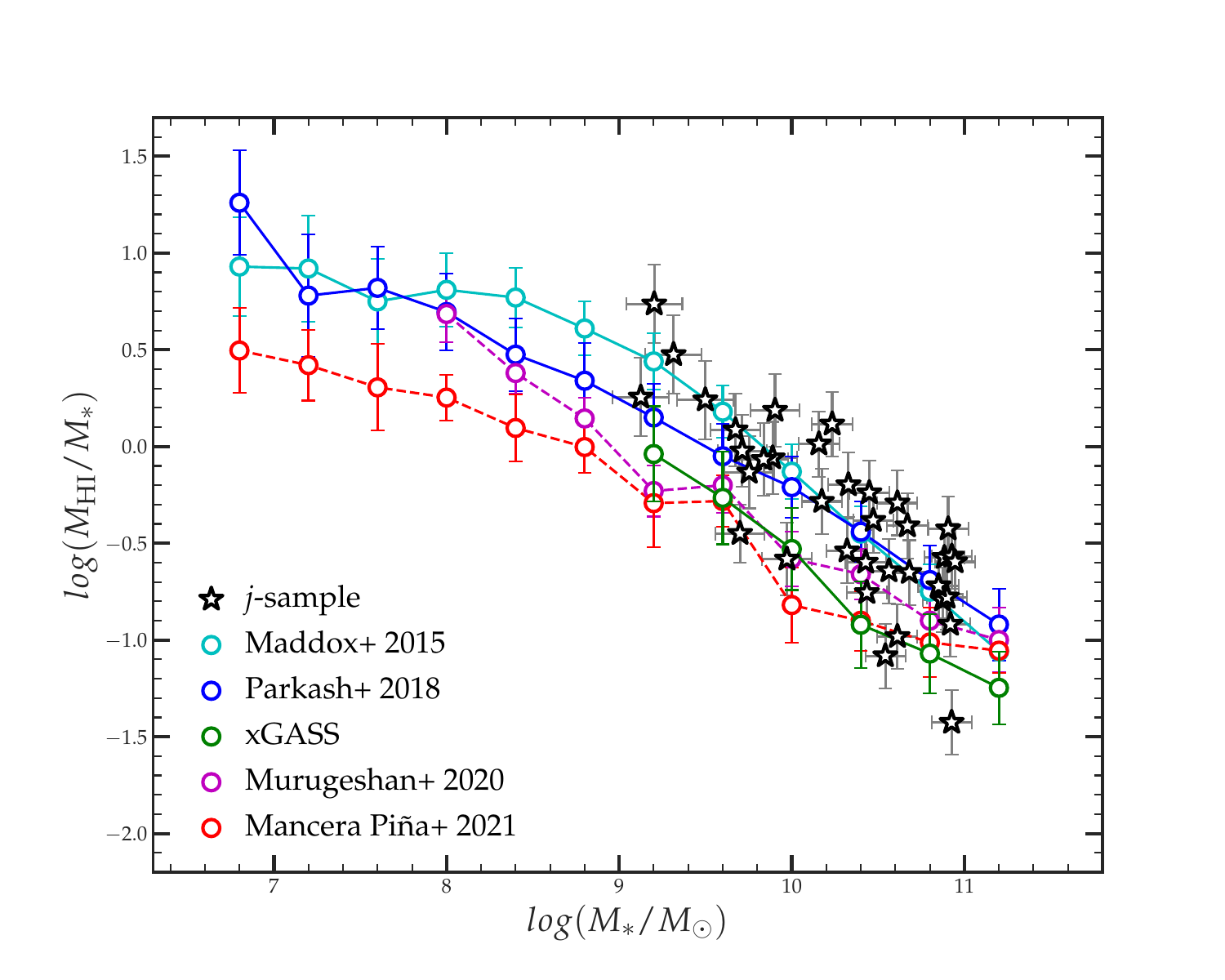}
    \vspace{-20pt}
    \caption{Comparison of the isolated $j$-sample's \hi-to-stellar mass fraction to existing samples in the literature.}\label{fig:mvsmfrac-lit}
\end{figure}

Compared to existing samples of non-isolated disc galaxies, the $j$-sample isolated galaxies are located in the high stellar mass end of the spectrum. This is shown in \Cref{fig:mvsmfrac-lit} where we compare the $j$-sample to the medians of other large galaxy samples in the literature: the \hi\ flux-limited ALFALFA-SDSS sample of 9153 galaxies \citep{Maddox2015} (cyan circles), the HICAT-WISE sample of 3158 galaxies \citep{Parkash2018} and the xGASS sample of 1179 galaxies \citep{Catinella2018}. We also include in the figure two \hi-selected, relatively smaller samples of resolved galaxies from \citepalias[114 galaxies;][]{Murugeshan2020} and \citepalias[157 galaxies;]{ManceraPina2021}, which we describe more extensively in \Cref{sec:jvsm}. The higher masses of the $j$-sample galaxies is caused by the velocity cut (threshold systemic velocity of 1500 \kms) imposed to isolated galaxies during the selection process, which systematically excludes low-mass galaxies.

\subsection{\HI\ Data}\label{sec:data:hi}
The measurement of the specific angular momentum requires good kinematic information of the candidate galaxies, i.e, reasonable spatial and spectral resolution data. Of the 587 galaxies making up the AMIGA {\it VerleyWISE} sample, we obtained good quality \hi\ data for 38 galaxies, compiled from various archival sources mainly obtained with the VLA, WSRT and GMRT telescopes. Particularly, eight galaxies were detected and retrieved from the first data release\footnote{Data available through https://vo.astron.nl} of the Apertif \citep[Aperture Tile In Focus;][]{Adams2022} survey.

The resolutions of the \hi\ data for each of the individual galaxies, as well as their noise levels and references are given in \Cref{tb:sources}. Eighteen of the 38 galaxies were published in the literature: we have obtained their reduced \hi\ datacubes (either through private communications or through the WHISP database\footnote{https://www.astro.rug.nl/~whisp/}), on which we performed the rotation curve modeling described in the last paragraph of this section. Additionally, data for 12 galaxies were retrieved from the VLA archive (their references are given in \Cref{tb:vla-refs}); for these, we proceeded to calibrate and image the data using a standard data reduction procedure\footnote{adapted from \url{https://github.com/AMIGA-IAA/hcg_hi_pipeline}} in CASA \citep[Common Astronomy Software Applications][]{McMullin2007}. Furthermore, data for 10 galaxies were retrieved from the Apertif data release, but those of CIG 468 and CIG 571 were discarded because the former lacked sufficient angular resolution and VLA data exist for the latter. For each of the remaining eight galaxies, we downloaded the spectral line data for the Apertif compound beam whose centre was closest to the galaxy of interest and which covered the correct frequency range, including the corresponding synthesized beam cube. The image cubes available in the archive are dirty cubes that have been output by the Apercal pipeline \citep{Adebahr2022}. We performed spline fitting on the dirty cubes along the spectral axis to remove any additional continuum residuals, and conducted automated source finding using \sofia\ \citep{Westmeier2021} to identify and mask emissions from the galaxy of interest. The data was then cleaned within the mask down to 0.5$\sigma$ using standard Miriad tools \citep{Sault1995}, and the clean cubes were primary beam corrected using the recommended Gaussian process regression models released with Apertif DR1\footnote{https://www.astron.nl/telescopes/wsrt-apertif/apertif-dr1-documentation/} \citep{Denes2022,Kutkin2022}. The properties of the \hi\ data for all galaxies in the $j$-sample are given in \Cref{tb:sources}, and their moment maps and position-velocity diagrams in \Cref{app:velfields}. Their physical resolutions range from 1.3 to 22.9 kpc, with 32 out of the 38 galaxies having synthesised beam sizes of ${<}10$ kpc. Furthermore, the column density sensitivities in the data range from $3\times10^{17}$ (for CIG 134) to ${\sim}1.5\times10^{20}\rm\,cm^{-2}$ (for CIG 676), estimated over a 20 \kms\ linewidth.
With the exception of the Apertif galaxies whose $3\sigma$ detection levels lie in the range ${\sim}1-3.5\rm\,M_\odot\,pc^{-2}$, the \hi\ in all galaxies in the sample is mapped to lower column density levels, reaching up to two orders of magnitude. This ensures that the full extent of the gas rotating with the discs is traced in most galaxies.

The \hi\ masses of a total of 844 AMIGA galaxies were measured in \citet{Jones2018} using single-dish data (GBT, Arecibo, Effelsberg and Nan\c{c}ay), including 587 galaxies of the Verley07b sample. All $j$-sample galaxies, except CIG 571, are comprised in these 587 galaxies. For this galaxy, we derived the \hi\ mass from the interferometric data cube and the optical distance. The downside of this method is the underestimation of the \hi\ mass since, by design, interferometers are poor at recovering the total \hi\ flux of galaxies. The \hi\ masses of the $j$-sample isolated galaxies cover the range $9.27 < \log(M_{\hi}/M_\odot) < 10.48$, with a median of $9.93\pm 0.05$.

From the \hi\ datacubes of the isolated galaxies in the $j$-sample, we made use of the \barolo\ package \citep{DiTeodoro2015} to model their rotation curve. The package takes as input the \hi\ cube of the galaxy, and performs a three-dimensional tilted-ring model fitting to determine the kinematic and geometrical parameters. An advantage of the three-dimensional (over the traditional two-dimensional) model-fitting, specifically with the \barolo\ package, is the minimisation of the beam smearing effects that arise when dealing with low-resolution data -- as is the case for some galaxies in our sample. For the algorithm to work efficiently one needs to provide initial guesses for the galaxy parameters; these are the kinematic centre, the systemic velocity, the line-of-sight inclination and position angle. For each galaxy in the $j$-sample, we took the optical parameters to be the initial parameters of the galaxy. To better improve the fitting procedure, we provide a 3D mask for each of the galaxies to \barolo. Each mask is constructed with the smooth and clip algorithm of \sofia\ at $4\sigma$, such that it essentially {\it only} contains the \hi\ emission of the corresponding galaxy. The output of \barolo\ comprises the \hi\ rotation curve and surface density profile of the galaxy, computed from concentric annuli, each characterised by a set of geometrical parameters (such as inclination and position angle) and centred on the kinematic centre of the galaxy. In \Cref{fig:rotpar} we show the variation of the geometric parameters with the radius, as well as the resulting surface density profiles. The values of the average fit results are given in \Cref{tb:kinpars}.

\begin{table*}
\caption{Properties of the \hi\ data for galaxies in the AMIGA angular momentum sample.The columns respectively list the CIG number, the NED name, the telescope used to observe the source, the size of the beam along the major and minor axes, the physical size of the beam along the major axis, the position angle of the beam, the velocity width of the cube, the $1\sigma$ noise of the data as well as the corresponding $3\sigma$ column density over a 20 \kms\ velocity width and the reference of the data.}\label{tb:sources}
\begin{tabular}{l l l c c c c c c l}
\hline
\hline
\multirow{2}{*}{CIG ID} & \multirow{2}{*}{Other Name} & \multirow{2}{*}{Tel.} & $\theta_{\rm maj} \times \theta_{\rm min}$ & $\theta_{\rm maj,\, kpc}$ & $\theta_{\rm PA}$ & $\Delta v$ & $\rm rms_{1\sigma}$ & $\rm\log{(N_{\sc HI,3\sigma}/cm^{-2})}$ & \multirow{2}{*}{Ref.} \\ 
 & & & arcsec. & $\rm kpc$ & deg. & $\rm km\,s^{-1}$ & $\rm mJy\,beam^{-1}$ & $\rm dex$ & \\ 
\hline
85 & UGC 01547 & GMRT & 22.6 $\times$ 18.8 & 3.9 & 24.2 & 13.4 & 0.62 & 20.0 & S12 \\ 
96 & NGC 0864 & VLA & 16.8 $\times$ 15.6 & 1.7 & -30.1 & 10.4 & 0.24 & 19.8 & RM18 \\ 
102 & UGC 01886 & WSRT & 29.5 $\times$ 24.5 & 9.5 & 0.0 & 17.0 & 0.05 & 18.6 & WHISP $^a$ \\ 
103 & NGC 0918 & VLA & 52.9 $\times$ 45.5 & 5.3 & 8.7 & 3.3 & 0.65 & 19.3 & VLA archive \\ 
123 & IC 0302 & VLA & 17.6 $\times$ 15.7 & 6.7 & -34.4 & 10.7 & 0.23 & 19.7 & VLA archive \\ 
134 & UGC 02883 & VLA & 68.1 $\times$ 52.5 & 22.6 & 24.6 & 10.7 & 0.05 & 17.9 & VLA archive \\ 
147 & NGC 1530 & WSRT & 33.1 $\times$ 23.5 & 5.8 & 0.0 & 16.8 & 0.04 & 18.6 & WHISP $^a$ \\ 
159 & UGC 03326 & WSRT & 29.7 $\times$ 23.0 & 8.2 & 0.0 & 16.9 & 0.04 & 18.6 & WHISP $^a$ \\ 
188 & UGC 03826 & GMRT & 43.1 $\times$ 36.6 & 5.1 & -22.1 & 3.5 & 1.10 & 19.7 & CS \\ 
232 & NGC 2532 & WSRT & 32.9 $\times$ 19.6 & 11.0 & 0.0 & 4.3 & 0.08 & 18.9 & WHISP $^a$ \\ 
240 & UGC 04326 & VLA & 62.1 $\times$ 54.4 & 19.4 & -56.8 & 10.6 & 0.51 & 19.0 & VLA archive \\ 
292 & NGC 2712 & VLA & 46.6 $\times$ 42.1 & 5.3 & -42.3 & 3.3 & 0.75 & 19.4 & P11 \\ 
314 & NGC 2776 & WSRT & 28.0 $\times$ 27.7 & 4.6 & 0.0 & 4.2 & 0.09 & 18.9 & WHISP $^a$ \\ 
329 & NGC 2862 & VLA & 13.9 $\times$ 13.5 & 3.5 & -22.8 & 21.2 & 0.17 & 19.8 & SG06 \\ 
359 & NGC 2960 & VLA & 65.9 $\times$ 60.8 & 19.6 & -12.42 & 20.6 & 0.27 & 18.6 & VLA archive \\ 
361 & NGC 2955 & VLA & 16.3 $\times$ 13.5 & 7.2 & -87.1 & 21.6 & 0.22 & 19.8 & SG06 \\ 
421 & UGC 05700 & VLA & 56.3 $\times$ 50.0 & 24.5 & 44.7 & 10.8 & 0.34 & 18.9 & E06 \\ 
463 & UGC 06162 & VLA & 45.6 $\times$ 42.5 & 6.3 & 75.1 & 10.5 & 0.33 & 19.1 & E06 \\ 
512 & UGC 06903 & GMRT & 17.3 $\times$ 13.3 & 1.7 & -49.6 & 13.9 & 0.47 & 20.1 & S12 \\ 
551 & UGC 07941 & VLA & 47.6 $\times$ 43.1 & 7.2 & 58.6 & 10.5 & 0.29 & 19.0 & E06 \\ 
553 & NGC 4719 & Apertif & 25.2 $\times$ 12.8 & 11.2 & 2.2 & 8.3 & 1.14 & 20.4 & Apertif $^b$ \\ 
571 & NGC 4964 & VLA & 67.2 $\times$ 52.3 & 10.8 & 18.1 & 20.9 & 0.76 & 19.2 & VLA archive \\ 
581 & NGC 5081 & Apertif & 30.8 $\times$ 14.5 & 12.8 & -2.3 & 8.3 & 1.14 & 20.2 & Apertif $^b$ \\ 
587 & UGC 08495 & Apertif & 25.1 $\times$ 17.9 & 12.3 & -3.1 & 8.3 & 1.25 & 20.3 & Apertif $^{b,c}$ \\ 
604 & NGC 5377 & WSRT & 33.2 $\times$ 28.7 & 3.7 & 0.0 & 8.3 & 0.02 & 18.1 & WHISP $^a$ \\ 
616 & UGC 09088 & VLA & 64.5 $\times$ 58.1 & 25.5 & -15.6 & 10.7 & 0.37 & 18.8 & VLA archive \\ 
626 & NGC 5584 & GMRT & 30.0 $\times$ 30.0 & 2.6 & 45.0 & 3.5 & 1.66 & 20.1 & P16 \\ 
660 & UGC 09730 & VLA & 58.8 $\times$ 46.3 & 8.6 & 66.6 & 10.5 & 0.67 & 19.2 & VLA archive \\ 
676 & UGC 09853 & Apertif & 17.3 $\times$ 13.2 & 6.6 & 0.0 & 8.3 & 1.51 & 20.6 & Apertif $^b$ \\ 
736 & NGC 6118 & VLA & 61.5 $\times$ 44.6 & 5.7 & -52.6 & 10.4 & 0.23 & 18.8 & VLA archive \\ 
744 & UGC 10437 & VLA & 67.2 $\times$ 55.7 & 11.5 & 74.7 & 10.5 & 0.31 & 18.7 & VLA archive \\ 
812 & NGC 6389 & VLA & 53.7 $\times$ 46.5 & 11.0 & 65.5 & 10.5 & 0.28 & 18.9 & E06 $^c$ \\ 
983 & UGC 12173 & Apertif & 24.5 $\times$ 13.6 & 7.9 & -1.7 & 8.3 & 1.08 & 20.3 & Apertif $^b$ \\ 
988 & UGC 12190 & Apertif & 29.7 $\times$ 13.4 & 14.2 & 1.5 & 8.3 & 1.32 & 20.3 & Apertif $^b$ \\ 
1000 & UGC 12260 & Apertif & 23.5 $\times$ 13.5 & 8.7 & -1.6 & 8.3 & 1.08 & 20.4 & Apertif $^b$ \\ 
1004 & NGC 7479 & VLA & 130.0 $\times$ 48.9 & 20.5 & 1.7 & 3.4 & 0.85 & 18.9 & VLA archive \\ 
1006 & UGC 12372 & Apertif & 25.0 $\times$ 14.0 & 9.1 & 0.5 & 8.3 & 0.76 & 20.2 & Apertif $^b$ \\ 
1019 & NGC 7664 & VLA & 56.4 $\times$ 47.9 & 13.1 & 8.7 & 3.4 & 0.68 & 19.2 & VLA archive \\ 
\hline
\end{tabular}\\
{\raggedright Notes:
$^a$ Data from the WHISP survey \citep{Swaters2002};
$^b$ Data from the Apertif DR1 \citep{Adams2022};
$^c$ Galaxy excluded from the isolated $j$-sample because of a non-detection in the \wise\ bands.}\\
{\raggedright\noindent References:
CS: courtesy of C. Sengupta;
E06: \citet{Espada2006}; P16: \citet{Ponomareva2016}; \citet{Portas2011}; RM18: \citet{Ramirez-Moreta2018}; S12: \citet{Sengupta2012}; SG06: \citet{Spekkens2006}}
\end{table*}

\begin{table}
\centering
\caption{References of the VLA archive data.}\label{tb:vla-refs}
\begin{tabular}{l c l l l}
\hline
\hline
CIG ID & VLA array & Project ID & Year & Project PI \\
\hline
103	& D	& AE175 & 2010 & L. Verdes-Montenegro \\
123	& C+D & AV276 & 2004 & L. Verdes-Montenegro \\
134	& D	& AV276 & 2004 & L. Verdes-Montenegro \\
240	& D	& AV276 & 2004 & L. Verdes-Montenegro \\
359	& D	& AG645 & 2003 & J. Greene \\
571	& D & AG645 & 2003 & J. Greene \\
616	& D	& AV276 & 2004 & L. Verdes-Montenegro \\
660	& D	& AV276 & 2004 & L. Verdes-Montenegro \\
736	& D	& AV276 & 2004 & L. Verdes-Montenegro \\
744	& D	& AV276 & 2004 & L. Verdes-Montenegro \\
1004 & D & AE175 & 2010 & L. Verdes-Montenegro \\
1019 & D & AE175 & 2010 & L. Verdes-Montenegro \\
\hline
\end{tabular}
\end{table}

\begin{table}
{\scriptsize
\caption{The global results of the \barolo\ fitting procedure. The R.A and Dec. columns represent the kinematic centre positions of the galaxies, $v_{\rm sys}$ their systemic velocities, and the last two columns respectively give their average inclinations and position angles.}\label{tb:kinpars}
\begin{tabular}{l l l l l c}
\hline
\hline
CIG ID & R.A (h:m:s) & Dec. (d:m:s) & $v_{\rm sys}$ (\kms) & Incl. (deg.) & P.A. (deg.) \\
\hline
85 & 02:03:21.0 & 22:02:31.1 & 2655.5 & 15.9 & 151.8 \\
96 & 02:15:27.1 & 06:00:19.9 & 1537.7 & 51.4 & 28.2 \\
102 & 02:26:01.8 & 39:28:19.5 & 4870.0 & 34.8 & 23.0 \\
103 & 02:25:50.9 & 18:29:47.4 & 1508.3 & 59.9 & 327.0 \\
123 & 03:12:51.3 & 04:42:30.0 & 5872.7 & 51.6 & 218.4 \\
134 & 03:52:14.1 & -01:30:29.0 & 5182.9 & 63.3 & 111.9 \\
147 & 04:23:27.5 & 75:17:58.7 & 2455.0 & 48.3 & 190.3 \\
159 & 05:32:09.0 & 77:17:00.0 & 4100.0 & 76.8 & 240.0 \\
188 & 07:24:28.6 & 61:41:38.0 & 1744.1 & 38.3 & 259.7 \\
232 & 08:10:15.1 & 33:57:16.7 & 5240.0 & 36.0 & 297.7 \\
240 & 08:20:35.2 & 68:36:01.0 & 4680.0 & 80.0 & 157.0 \\
292 & 08:59:30.6 & 44:54:35.0 & 1870.0 & 77.2 & 4.8 \\
314 & 09:12:15.1 & 44:57:09.3 & 2615.8 & 39.0 & 306.7 \\
329 & 09:24:55.2 & 26:46:25.0 & 4086.0 & 79.0 & 292.6 \\
359 & 09:40:35.7 & 03:34:37.0 & 4899.5 & 46.4 & 224.8 \\
361 & 09:41:16.8 & 35:52:58.1 & 7015.0 & 63.8 & 169.5 \\
421 & 10:31:15.1 & 72:07:35.0 & 6652.0 & 39.1 & 18.6 \\
463 & 11:06:54.6 & 51:12:12.1 & 2212.7 & 68.8 & 88.2 \\
512 & 11:55:37.1 & 01:14:14.1 & 1897.3 & 32.1 & 133.0 \\
551 & 12:46:00.7 & 64:34:21.8 & 2306.8 & 68.7 & 8.4 \\
553 & 12:50:08.9 & 33:09:23.2 & 7056.2 & 25.6 & 47.7 \\
571 & 13:05:26.1 & 56:19:29.0 & 2544.7 & 56.2 & 320.1 \\
581 & 13:19:08.3 & 28:30:29.1 & 6601.2 & 75.0 & 99.9 \\
587 & 13:29:56.6 & 50:52:52.1 & 7618.9 & 55.7 & 45.6 \\
604 & 13:56:16.2 & 47:14:14.5 & 1799.6 & 65.2 & 210.9 \\
616 & 03:12:50.3 & 04:42:26.0 & 5873.6 & 43.5 & 222.6 \\
626 & 14:22:23.4 & -00:23:25.6 & 1618.4 & 47.5 & 150.9 \\
660 & 15:03:56.8 & 77:38:18.0 & 2136.7 & 44.6 & 45.2 \\
676 & 15:25:47.1 & 52:26:43.9 & 5817.5 & 75.0 & 271.3 \\
736 & 16:21:48.0 & -02:17:03.0 & 1596.1 & 70.3 & 47.7 \\
744 & 16:31:07.0 & 43:20:47.5 & 2614.8 & 33.0 & 349.0 \\
812 & 17:32:39.1 & 16:24:06.0 & 3130.0 & 39.9 & 311.2 \\
983 & 22:43:51.8 & 38:22:40.6 & 4712.3 & 62.0 & 257.2 \\
988 & 22:48:06.6 & 28:17:36.0 & 7241.1 & 82.7 & 354.0 \\
1000 & 22:56:32.1 & 37:44:21.3 & 5537.6 & 75.8 & 30.5 \\
1004 & 23:04:57.2 & 12:19:13.7 & 2377.9 & 46.6 & 211.6 \\
1006 & 23:07:01.0 & 35:46:33.7 & 5454.9 & 43.0 & 30.1 \\
1019 & 23:26:40.0 & 25:04:51.4 & 3480.4 & 54.9 & 87.4 \\
\hline
\end{tabular}}
\end{table}

\subsection{Mid-Infrared Data}\label{sec:data:wise}
We use mid-infrared \wise\ \citep{Wright2010} observations to trace the stellar components of the AMIGA galaxies. More specifically, we refer to the \wise\ Extended Source Catalog \citep[WXSC,][]{Jarrett2019} to obtain the photometric data of the AMIGA galaxies: these include the {\it W1} ($3.4\um$) and {\it W2} ($4.6\um$) fluxes -- sensitive to stellar populations -- of the galaxies, the stellar surface brightness profiles and the $W1$-$W2$ colours. The full source characterisation, including the star-formation sensitive bands at $12\um$ ($W3$) and $23\um$ ($W4$), are available in \citet{Jarrett2023}.

The \wise\ photometries of the AMIGA galaxies were derived following the method described in \citet{Parkash2018} and \citet{Jarrett2013,Jarrett2019}; first, image mosaics were constructed from single native \wise\ frames using a technique detailed in \citet{Jarrett2012}, and resampled to a $1"$ pixel scale -- relative to the beam. Because of the modest angular size of the AMIGA galaxies (their optical radii range from $10.8''$ to $4.6'$), the above pixel scale was appropriate to accommodate their angular sizes and no extra processing step was needed as is the case for some large nearby objects processed in \citet{Jarrett2019}. 

Of the 791 galaxies in the Verley07b sample, infrared photometries of 632 galaxies were successfully and reliably extracted from the WSXC catalogue. However, only 587 of those also happen to have \hi\ masses available. For each of those, the total flux was measured in each of the four \wise\, bands -- including the {\it W3} ($12\um$) and {\it W4} ($23\um$) bands. The {\it W1} and {\it W2} total fluxes were estimated using a technique developed for the 2MASS \citep{Jarrett2000}, which consists of fitting a double S\'ersic profile to the axisymmetric radial flux distribution. This way, both the star-forming disc and bulge components are each represented by a single S\'ersic profile. Owing to the lower sensitivity of the longer wavelength bands {\it W3} and {\it W4}, the total fluxes of part of the sample galaxies in these bands are obtained through extrapolation of their extent to three disc scale lengths after fitting their light profiles with the double S\'ersic function. However, since these longer wavelength fluxes are not used in this work, it is not relevant to discuss their measurements here. For a full description and discussion of their derivation, we refer the reader to \citet{Jarrett2019}.

Besides the total flux, the global stellar mass was also estimated for each of the \wise\ detections. This was done by estimating the mass-to-light ratio $M/L_{W1}$ in the {\it W1} band from the $W1-W2$ colour, and converting the {\it W1} flux density to the luminosity $L_{W1}$. As specified in \citet{Jarrett2019}, this is based on the assumption that the observed {\it W1} light is emitted by the galaxy's sole stellar population, and that the post-AGBs population are not significantly contributing to the near-infrared brightness. To evaluate $M/L_{W1}$, we make use of the new GAMA colour-to-mass calibration method in \citet{Jarrett2023}. The average $M/L_{W1}$ found therein is $0.35\pm0.05$, about 30\% lower than the mass-to-light ratio value of of 0.5 (in the $3.6\rm\mu m$ band) adopted in \citetalias{ManceraPina2021} for disc-dominated galaxies. As for \citetalias{Murugeshan2020}, the authors estimated their stellar masses from $K_{\rm s}$ magnitudes based on the calibration from \citet{Wen2013}.

Additionally to these parameters, we have also measured the {\it W1} and {\it W2} light profiles -- the surface brightness at different radii -- of a subset of 449 galaxies, including the 36 isolated galaxies in the $j$-sample (except CIG 587 \& 812). These light profiles, presented in \Cref{fig:rotpar}, provide information on the distribution of the stellar density as a function of the radius, necessary for measuring the stellar specific angular momentum (see \Cref{sec:j} below).

\section{The Specific Angular Momentum}\label{sec:j}
The specific angular momentum of a disc galaxy is defined as $j \equiv J/M$, where $J$ is the orbital angular momentum of the galaxy and $M$ its total mass. More explicitly, the specific angular momentum carried by a galaxy's component $i$ of radius $R$ can be written as
\begin{equation}\label{eq:jk}
    j_i(<R) = \frac{\int_0^R r^2\, \Sigma_i(r)\, v_i(r)\, dr}{\int_0^R r\, \Sigma_i(r)\, dr},
\end{equation}
where $\Sigma_i(r)$ and $v_i(r)$ are respectively the surface density and velocity of the component $i$ at radius $r$. The errors associated with $j_i$ are estimated following \citet{Posti2018} and approximating the disc scale length $R_{\rm d}$ to ${\sim}30\%$ the radius at the 25th magnitude $R_{25}$ (e.g, \citealt{{Korsaga2018}} find $R_{\rm d}\sim0.35R_{25}$):
\begin{equation}\label{eq:djk}
    \delta j_i = 0.3 R_{25} \sqrt{\frac{1}{N}\sum_{n}^{N}\delta_{v_n}^2 + \left(\frac{V_{\rm flat}}{\tan{(\rm incl.)}}\delta_{\rm incl.}\right)^2 + \left(V_{\rm flat}\frac{\delta_D}{D}\right)^2},
\end{equation}
where the distance $D$, inclination {\it incl.} and radius $R_{25}$ are taken from \citet{Lisenfeld2011}, and the flat velocity $V_{\rm flat}$ evaluated from the rotation curve (see \Cref{sec:disc:btf}). For all galaxies, we assume a ${\sim}20\%$ uncertainty on the distance (for reference, \citealt{Posti2018} find the distance errors of the SPARC galaxies to fluctuate between $10-30\%$); furthermore, the error associated to the inclination is taken to be the difference between the inclinations of the \hi\ and stellar discs. Finally, the error $\delta_{v_n}$ associated to the rotation velocity is estimated at each point $n$ of the rotation curve, and $N$ represents the number of radii at which $j_i$ is evaluated. We note that, since \Cref{eq:djk} uses the optical disc scale length for both the stellar and gas components, and given that the \hi\ usually extends further than the stars in disc galaxies \citep[e.g.,][]{Broeils1997}, $\delta j_{\rm gas}$ could somewhat be underestimated. As such, it must be regarded only as an indication of the uncertainties on \jgas.
Furthermore, we consider that the baryonic mass of a galaxy is distributed among its two major constituents: the stellar and gas components. In the following, we denote the specific angular momenta of these two components as $j_\star$ and $j_{\rm gas}$, respectively. Therefore, the total baryonic angular momentum can be expressed as
\begin{equation}\label{eq:jbar}
    j_{\rm bar} = f_{\rm gas}\,j_{\rm gas} + (1-f_{\rm gas}) j_\star
\end{equation}
where $f_{\rm gas} = M_{\rm gas}/(M_{\rm gas}+M_\star)$ denotes the galaxy's gas fraction.

The gas surface densities in \Cref{eq:jk} are obtained by applying a factor of 1.35 to the \hi\ surface densities (i.e, $\Sigma_{\rm gas} = 1.35\,\Sigma_{\hi}$) to account for the helium. We ignore the molecular component of the gas since no CO observations could be found for the galaxies. Also, the contribution of the molecular gas to the baryonic angular momentum is expected to be negligible based on previous studies \citep[see e.g.,][in their appendix]{ManceraPina2021a}. We compute $j_{\rm gas}$ by simply substituting the \hi\ rotation velocities and the gas surface densities in \Cref{eq:jk}.
Because of the difficulty associated with correctly determining the velocities of the stars, we approximate these to the gas velocities -- i.e, $v_\star(r) \equiv v_{\rm gas}(r)$ -- and therefore determine $j_\star$ using \Cref{eq:jk} with the stellar surface densities derived from the \wise\ $3.4\um$ band photometry (see \citealt{Sorgho2019} for how the $3.4\um$ photometry is used to trace the kinematics of the stellar disc). This approximation holds for massive disc galaxies whose stellar components exhibit regular rotational motions, unlike dwarf galaxies in which random, non-circular motions are significant. On the other hand, since the AMIGA galaxies were selected to have velocities greater than 1500 \kms, very few low-mass galaxies were included in the sample. Specifically for the $j$-sample, \Cref{fig:mvsmfrac-lit} shows that all 36 galaxies have stellar masses higher than $10^9\,M_{\odot}$, which makes the approximation suited for the present study. 

\subsection{The specific angular momentum of the atomic and stellar discs}

Mathematically, the specific angular momentum is a combined measure of how large a galaxy is and how fast it rotates. Therefore, large and fast-rotating galaxies are expected to possess a higher specific angular momentum than small, slow-rotating galaxies. On the other hand, early-type spirals are known to be larger and have higher circular velocities than their late-type counterparts, which in turn rotate faster than irregular galaxies.

\begin{figure}
    \centering
    \includegraphics[width=\columnwidth]{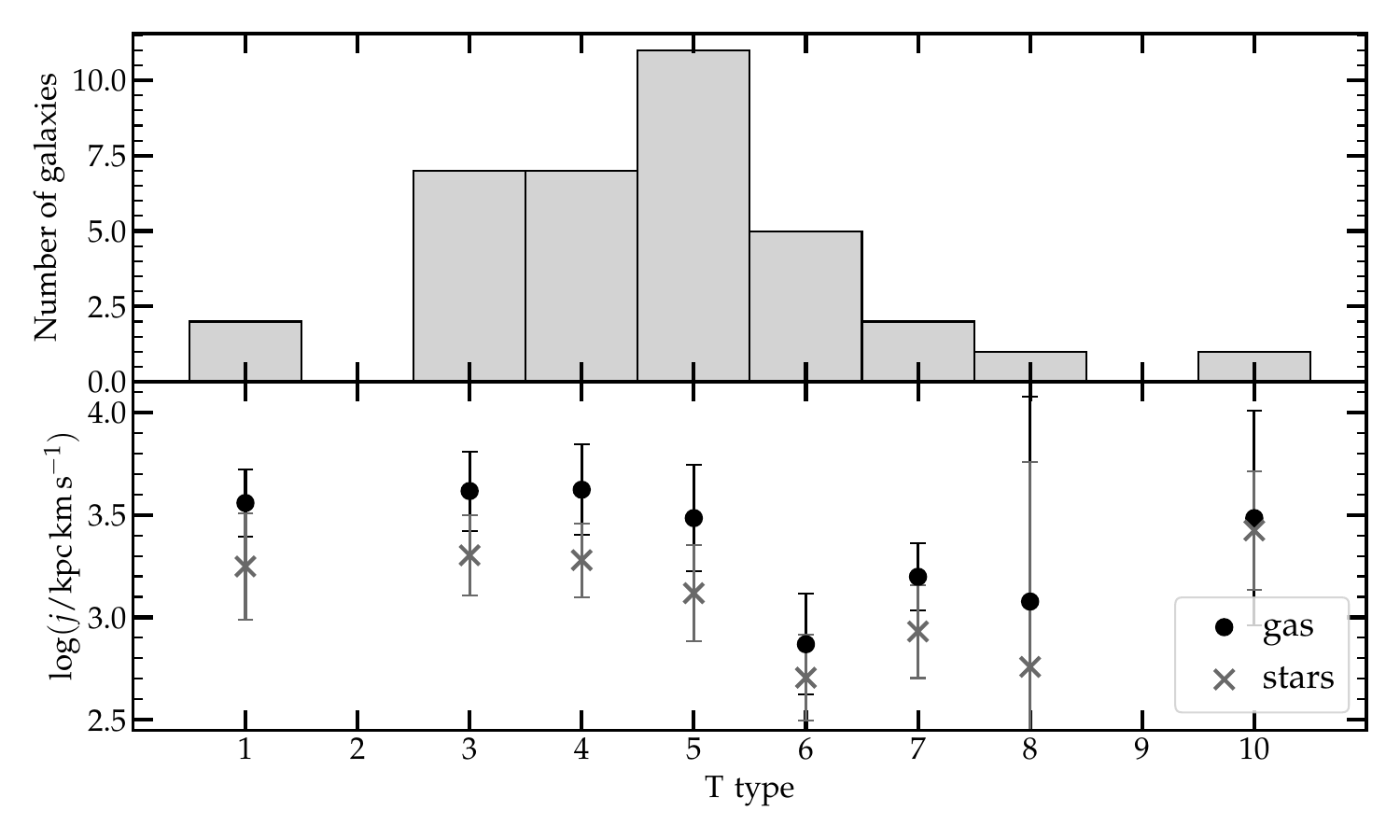}
    \vspace{-20pt}
    \caption{Angular momentum and morphologies of the galaxies in the isolated $j$-sample (including CIG 587 \& 812). {\it Top}: morphological distribution of the sample; {\it bottom}: the specific angular momentum as a function of the morphological type, for the atomic gas and stellar discs.}\label{fig:jt}
\end{figure}

\begin{figure}
    \centering
    \includegraphics[width=\columnwidth]{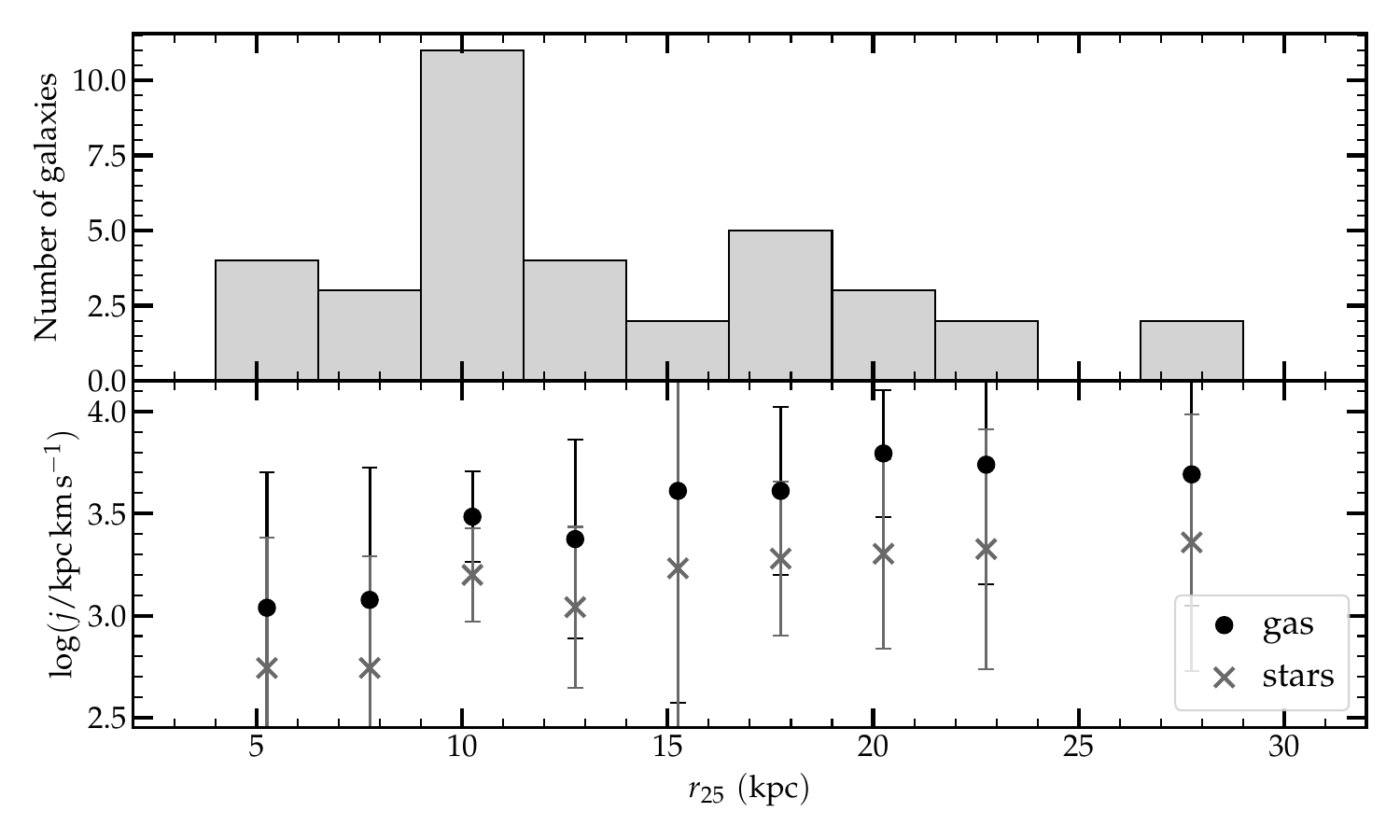}
    \vspace{-20pt}
    \caption{Same as \Cref{fig:jt}, but as a function of the optical radius.}\label{fig:jr}
\end{figure}

The isolated $j$-sample is constituted of 36 galaxies of mostly late morphological types (Sa to Irr), dominated by Sb and Sc morphologies (see top panel of \Cref{fig:jt}). For each of the atomic gas and stellar components of the galaxies in the sample, we show in the bottom panel of the figure the median specific angular momentum plotted as a function of the morphological type T. The T morphologies are referenced from the RC3 scale \citep{DeVaucouleurs1991}, where T values increase from early to late-type morphologies, such that T=0 corresponds to an S0a type and T=10 indicates an Irr galaxy.
As expected, the angular momentum is highest for early-type spirals and decreases towards the late types, until about $\rm T \approx 6-7$. The mean $j$ values at the later morphological types (T = 8 \& 10) increase, but since they only contain one galaxy each, it is not clear what the actual trend is at these morphologies. A reverse correlation is seen when the specific angular momentum is plotted against the optical radius (B-band isophotal radius at the 25th magnitude taken from \citealt{FernandezLorenzo2012}), as seen in \Cref{fig:jr}.
As expected, \jgas\ is systematically higher than \jstar; this is because, on average, the gas is distributed at larger radii than the stars \citep[e.g.,][]{Broeils1997,Swaters2002}, and is therefore expected to carry more angular momentum.


\section{The Specific Angular Momentum -- Mass Relation}\label{sec:jvsm}
The current galaxy formation paradigm predicts that both the dark matter halo and baryonic disc acquire their angular momentum through gravitational torques, during the proto-galaxy formation phase \citep[e.g.,][]{Peebles1969,White1984}. The resulting disc, formed via the collapse and condensation of cold gas within the potential wells of the parent halo, ends up with the same specific angular momentum as the halo \citep[e.g.,][although the latter predicts the possibility that a fraction of the initial mass and angular momentum will not settle into the disc]{Fall1980,Mo1998}. As a result, it should be expected that the baryonic $j$ behaves as $j_{\rm bar} \propto M_{\rm bar}^{2/3}$, similarly to the DM halo. However, current observations are not consistent with this prediction. As pointed by some studies, not all the baryons carrying angular momentum may condense into the galaxy disc, explaining the discrepancy with the expectation \citep[e.g,][]{Kassin2012}. Furthermore, with numerical simulations becoming more available, it has become evident that more mechanisms are at play in the angular momentum acquisition and conservation of discs throughout their lifetime; for example, the different interactions that galaxies undergo with their environment, such as mergers, are capable of affecting their total baryonic angular momentum \citep[e.g.,][]{Danovich2015,Lagos2017,Jiang2019}. In this section, we investigate the Fall relation (\jbar\ vs. \Mbar) for the isolated galaxies in the $j$-sample and perform a comparison with the samples of non-isolated galaxies.

\subsection{The comparison samples} \label{sec:jvsm:local}
To investigate whether the angular momentum of isolated galaxies behave in a particular way, in the context of galaxy evolution, we compare the AMIGA galaxies to samples of non-isolated galaxies\footnote{Since the isolation parameters $\eta_k$ and $Q$ were not determined for the galaxies making up these samples, we only consider them non-isolated in a statistical sense: the samples may include a few isolated galaxies, but the majority are not more isolated than field galaxies.} found in the literature: the large two samples \citetalias{Murugeshan2020} (114 galaxies) and \citetalias{ManceraPina2021} (157 galaxies) mentioned above, and three moderately small samples from \citet[][14 galaxies]{Butler2017}, \citet[][11 galaxies]{Kurapati2018} and \citet[][16 galaxies]{Kurapati2021}. 
The specific angular momentum as well as mass values are taken from the corresponding studies, which use somewhat similar methods to determine the gas kinematics. \citetalias{Murugeshan2020}, \citetalias{ManceraPina2021} and \citet{Kurapati2018} derive their rotation curves similarly to the method used in the present work, with the difference that \citet{Kurapati2018} use the FAT \citep[Fully Automated TiRiFiC;][]{Kamphuis2015} package instead of \barolo. Additionally, \citetalias{ManceraPina2021} add an asymmetric drift term to their rotational velocities to correct for the non-circular motions typically prominent in low-mass galaxies. \citet{Butler2017} and \citet[][who use rotation curves from \citealt{Verheijen2001a}]{Kurapati2021} build their rotation curves by fitting a tilted-ring model onto concentric ellipses taken along the spatial extent of the \hi\ discs, with the assumption that the rotation curve has a parametric functional form.

The 114 galaxies in \citetalias{Murugeshan2020} were selected from the Westerbork \hi\ Survey of Spiral and Irregular Galaxies \citep[WHISP;][]{Swaters2002}, such that their \hi\ radius spans at least five resolution elements in the $30''$ resolution data, and their inclination between $20\dg$ and $80\dg$. The sample contains a mix of low, intermediate and high mass galaxies, with \hi\ and stellar masses spanning about three  and five orders of magnitude, respectively ($7.8 < \log({M_\HI}/M_\odot) < 10.5$ \& $6.7 < \log(M_{\rm star}/M_\odot) < 11.5$). The stellar component of each of the individual galaxies was traced using 2MASS \citep[Two Micron All-Sky Survey;][]{Skrutskie2006} $\rm K_s$-band photometries (see \citetalias{Murugeshan2020}).
Similarly, the \citetalias{ManceraPina2021} sample was constructed by compiling 157 galaxies from six main sources: 90 spirals from the SPARC catalogue \citep[{\it Spitzer} Photometry and Accurate Rotation Curves;][]{Lelli2016}, 30 from a sample of spirals by \citet{Ponomareva2016}, 16 dwarfs from the LITTLE-THINGS sample \citep[Local Irregulars That Trace Luminosity Extremes, The \hi\ Nearby Galaxy Survey;][]{Hunter2012}, 14 dwarfs from the LVHIS sample \citep[Local Volume \hi\ Survey;][]{Koribalski2018}, four dwarfs from the VLA-ANGST sample \citep[Very Large Array-ACS Nearby Galaxy Survey Treasury;][]{Ott2012} and finally three dwarfs from the WHISP sample. To derive the properties of the galaxies' stellar components, the authors made use of either the {\it Spitzer} $\rm 3.6\mu m$ or the {\it H}-band $\rm 1.65\mu m$ photometry.

Of the above two comparison samples, one (CIG 626) and five (CIG 102, 147, 232, 314 \& 604) galaxies respectively from the \citetalias{ManceraPina2021} and \citetalias{Murugeshan2020} samples are included in the $j$-sample. In fact, 17 galaxies in each of these two samples are catalogued in the initial CIG sample of isolated galaxies \citep{Karachentseva1973}, but were discarded from the sample of 950 ``high-velocity'' galaxies discussed in \Cref{sec:data:amiga} because their systemic velocities are $v<1500$ \kms. These six galaxies will be discarded from the two samples in the analysis follows. Furthermore, five of the remaining 156 galaxies of \citetalias{ManceraPina2021} did not have available \jstar\ values, these were therefore removed from the sample. The final sizes of the samples are thus \sizeman{} and \sizemur{} galaxies, respectively for \citetalias{ManceraPina2021} and \citetalias{Murugeshan2020}.

Unlike the previous two samples, the last three have sizes about an order of magnitude smaller. The \citet{Kurapati2021} sample, containing 16 {\it normal}, regularly rotating spiral galaxies, was originally drawn from \citet{Verheijen2001}'s sample of galaxies in the Ursa Major region. The stellar component of each of the galaxies in the sample was derived using {\it K}-band luminosity profiles. The baryonic masses of the galaxies in the sample range from $9.25 < \log({M_{\rm bar}}/M_\odot) < 11$, with only UGC 7089 having a baryonic mass lower than $10^{9.6}\,\rm M_\odot$ (corresponding to the $j$-sample's lower limit, see \Cref{sec:jvsm:mass-cut}).

On the other hand, the \citet{Kurapati2018} and \citet{Butler2017} samples are essentially made of dwarf galaxies respectively selected from the nearby Lynx-Cancer void \citep{Pustilnik2011} and the LITTLE-THINGS sample. \citet{Kurapati2018} made use of SDSS \citep{Ahn2012} and PanSTARRS \citep{Flewelling2016} $g$-band luminosities and $g-i$ colours to trace the stellar components of the galaxies, while those of the \citep{Butler2017} sample were obtained from {\it Spitzer} $\rm 3.6\mu m$ images. It should be noted that none of \citet{Butler2017}, \citet{Kurapati2018} or \citet{Kurapati2021} samples include galaxies from the $j$-sample.

\subsection{The Fall relation: isolated vs. non-isolated galaxies} \label{sec:jvsm:isol}
In \Cref{fig:jvsm-bar} we present the total baryonic angular momentum \jbar\ of the AMIGA galaxies, along with a comparison with the non-isolated samples mentioned above: the larger \citetalias{Murugeshan2020} and \citetalias{ManceraPina2021} samples, and the three smaller samples from \citet{Butler2017}, \citet{Kurapati2018} and \citet{Kurapati2021}. The \citet{Butler2017} sample includes the galaxy UGC 8508, which the authors found to be an outlier in the mass-$j$ relation because of its abnormally high \jbar\ for its modest baryonic mass. Therefore, we accordingly remove UGC 8508 from the angular momentum analyses that follow. The left panel of the figure shows that the galaxies in the AMIGA angular momentum sample have \jbar\ values that are similar to those of non-isolated galaxies, with the noticeable difference that they occupy the upper end of the parameter space. A linear regression of the form
\begin{equation}\label{eq:j_eq}
    \log{(j_{\rm bar}/{\rm kpc\,km\,s^{-1}})} = \alpha \left[\log{(M_{\rm bar}/M_\odot)} - 10\right] + c
\end{equation}
was fit to the angular momentum sample and to the two largest samples of non-isolated galaxies using Bayesian inference, specifically a \verb+Python+ implementation of the Monte Carlo Markov Chain (MCMC) in the open-source {\sc PyMC3}\footnote{Documentation at https://docs.pymc.io/en/v3/index.html} package \citep{Salvatier2016}. The fitting procedure consists of assuming priors for three parameters: the slope $\alpha$, the intercept $c$ and the intrinsic scatter $\sigma$. For the slope and intercept, a gaussian prior with a mean of respectively 1 and 2 and a standard deviation of 4 was used, while for the scatter we chose an exponential prior of coefficient 1.
Next, instead of a gaussian distribution, we adopt a Student-$t$ distribution (with a degree of freedom $\nu$ for which a half-normal distribution of standard deviation 5 was chosen as prior, see \Cref{app:posterior}) to explore the likelihood. Because of its {\it fatter} tails, the $t$-distribution has the added advantage of minimising the influence of the outliers. Given the modest size of the samples in this study, especially the AMIGA $j$-sample of 36 galaxies, this distribution proved to be more effective at constraining the free parameters.

We obtain a best-fit slope of $\alphaj$ for the isolated $j$-sample, about 20\% lower than the theoretical slope of $\sim 2/3$ predicted in hierarchical models for dark matter (we discuss this in \Cref{sec:disc:cdm}). Since we have altered the \citetalias{ManceraPina2021} and \citetalias{Murugeshan2020} samples, and for consistency, we re-perform linear regression fits on these. As a reminder to the reader, the main changes in the samples are (i) the removal of galaxies that overlap with the $j$-sample and the inclusion of galaxies previously discarded by \citetalias{ManceraPina2021}, whose \jbar\ values are non-converging.

The re-derived best fit slopes are $\alphaman$ and $\alphamur$, respectively for the \citetalias{ManceraPina2021} and \citetalias{Murugeshan2020} samples. For context, the best-fit values of the slope obtained in the previous studies are $0.60\pm0.02$ and $0.55\pm0.02$, respectively for the original \citetalias{ManceraPina2021} and \citetalias{Murugeshan2020} samples. As a sanity check, we performed the fit on these original, non-altered samples and found consistent results with the original studies. It should be noted that the authors used a fitting method different than what we adopted here: both \citetalias{Murugeshan2020} and \citetalias{ManceraPina2021} performed the fit with the {\sc hyper-fit} package \citep{Robotham2015}, a tool designed for fitting linear models to data with multivariate gaussian uncertainties.

The results of the linear regressions are summarised in \Cref{tb:fit-amiga}. The first two columns of the table show respectively the different samples and their sizes, while the last three columns list respectively the slope $\alpha$, intercept $c$ and intrinsic scatter $\sigma$ obtained from fitting \Cref{eq:j_eq} to each of the samples.

\begin{table}
    \centering
    \caption{The results of linear regressions to the different samples.}
    \label{tb:fit-amiga}
    \begin{tabular}{lcccc}
    \hline
    \hline
    Sample & size & $\alpha$ & $c$ & $\sigma$ \\
    \hline
    AMIGA & 36 & $\alphaj$ & $\cj$ & $\sigmaj$\\
    M20 & \sizemur & $\alphamur$ & $\cmur$ & $\sigmamur$ \\
    MP21 & \sizeman & $\alphaman$ & $\cman$ & $\sigmaman$ \\
    \hline
    \end{tabular}
\end{table}

As mentioned in \Cref{sec:data:amiga}, the isolation criteria adopted in the present analysis are those defined in \citet{Verley2007b}, which were applied on a larger galaxy sample than the study conducted in \citet{Argudo-Fernandez2013} because of the limited SDSS footprint. In fact, \citet{Argudo-Fernandez2013} accounted for the spectroscopic redshift when evaluating the galaxies' isolation parameters, which is not available for a significant subset of the sample. This results in a very strict definition of isolation, given that the AMIGA galaxies were selected from a previously-built catalogue of isolated galaxies \citep{Karachentseva1973,Verdes-Montenegro2005a}. A cross-match between the $j$-sample and the sample considered in \citet{Argudo-Fernandez2013} results in only 16 galaxies, of which one galaxy (CIG 361) does not meet the isolation criteria. For the sake of a fair comparison, we highlight this stricter sample to the right panel of \Cref{fig:jvsm-bar} (circled stars and grey dash-dotted line). The slope measured for these galaxies is lower than that of the $j$-sample, but the uncertainty associated with the fit results, as well as the lack of systematic offset between the lines of best fit, suggests that they do not substantially differ from the $j$-sample.

\subsection{The Fall relation: low-mass vs. high-mass galaxies} \label{sec:jvsm:mass-cut}
\begin{figure*}
    \centering
    \begin{minipage}{0.49\textwidth}
        \centering
        \includegraphics[width=\textwidth]{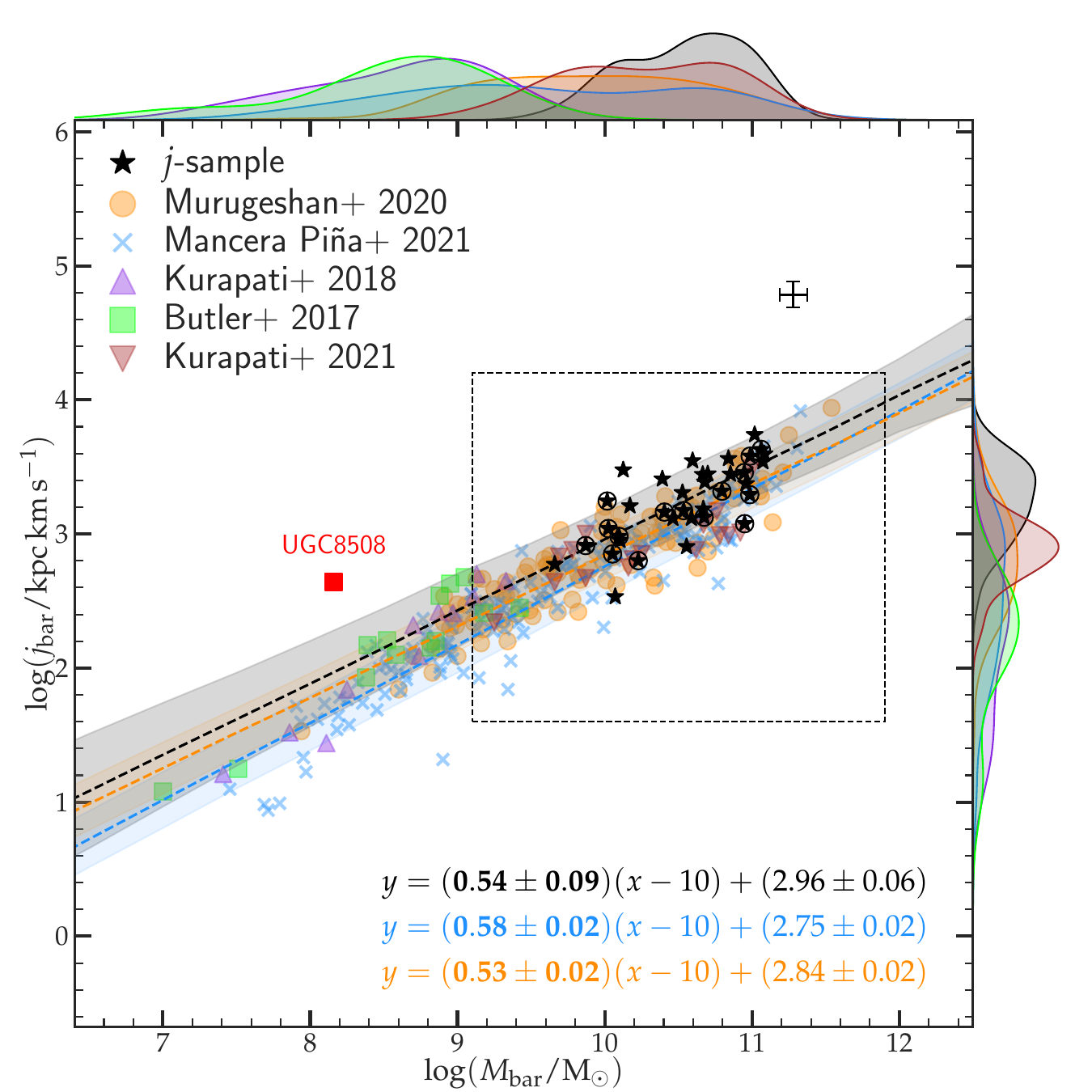}
    \end{minipage}
    \begin{minipage}{0.49\textwidth}
        \centering
        \includegraphics[width=\textwidth]{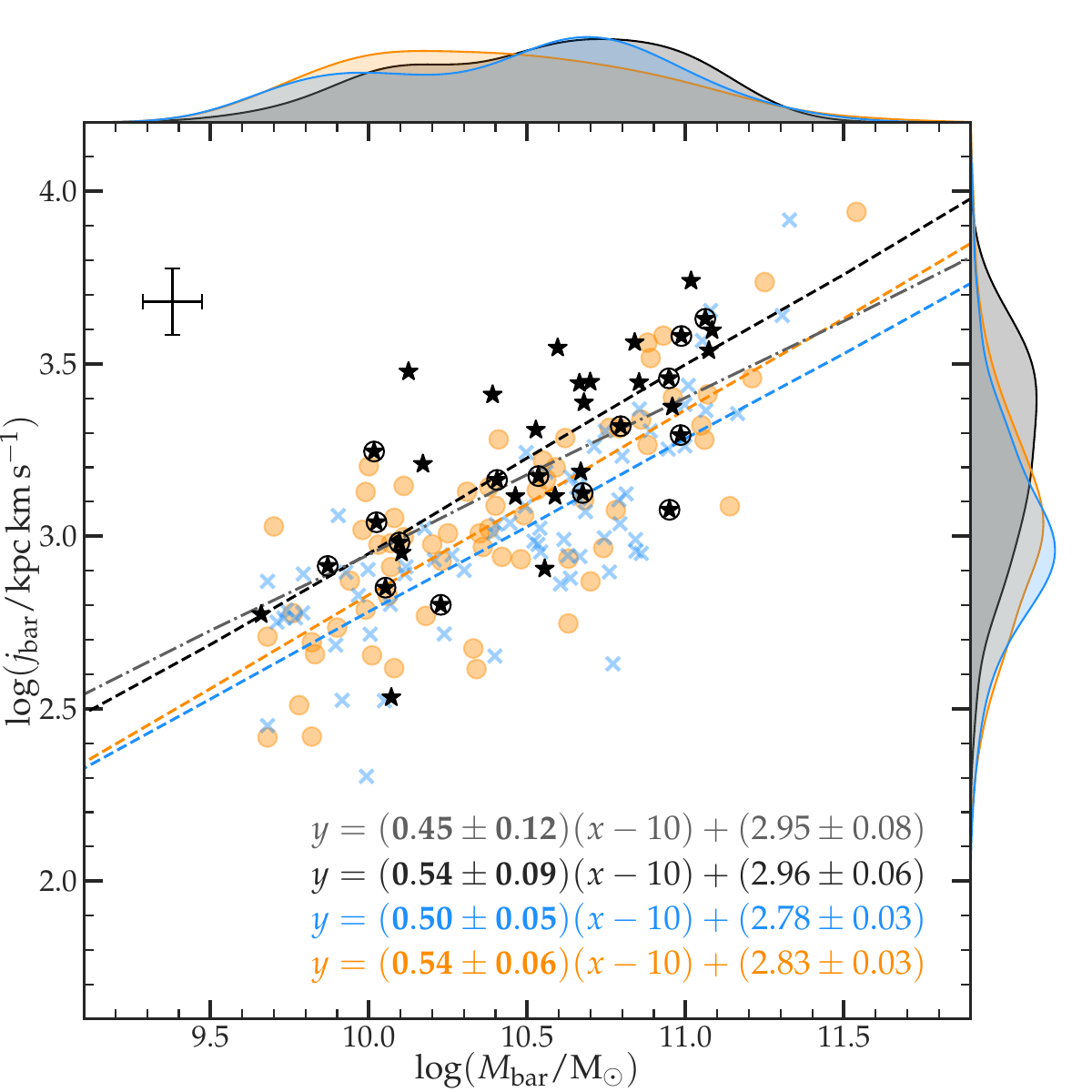}
    \end{minipage}
    \caption{Specific angular momentum as a function of the baryonic mass for the different samples. {\it Left:} all samples are shown across the total baryonic mass range, with the linear regressions for the isolated $j$-sample and the \citetalias{Murugeshan2020} and \citetalias{ManceraPina2021} samples shown by dashed lines. The dashed square shows the limits of the right panel, and the red filled square denotes the outlier galaxy UGC 8508. {\it Right:} only galaxies in the \citetalias{Murugeshan2020} and \citetalias{ManceraPina2021} samples for which $M_{\rm bar} > 10^{9.6}\,M_\odot$ are considered, and the resulting linear regressions are shown by dashed lines. The circled stars and their grey line of best fit represent the galaxies included in the \citet{Argudo-Fernandez2013} sample (see \Cref{sec:jvsm:isol}) The typical error bars are shown in each of the panels.} \label{fig:jvsm-bar}
\end{figure*}

Could the narrower mass range of the AMIGA sample induce discrepancies into the results of the regressions? To probe this, we applied a lower cut of $\log(M_{\rm bar}/M_\odot) = 9.6$ -- corresponding to the lower mass limit of the isolated $j$-sample -- on the baryonic mass of the \citetalias{Murugeshan2020} and \citetalias{ManceraPina2021} samples. This resulted in \sizemurhigh\ and \sizemanhigh\ galaxies, respectively for the \citetalias{Murugeshan2020} and \citetalias{ManceraPina2021} samples (see \Cref{tb:fits}). A linear regression fit on these new, high-mass samples is shown in the right panel of \Cref{fig:jvsm-bar}: while the slope of the \citetalias{Murugeshan2020} sample has almost remained constant, that of the \citetalias{ManceraPina2021} sample has decreased from $\alphaman$ to $\alphamanhigh$. Overall, the best-fit lines of these samples remain below that of the $j$-sample. This is further seen in the distributions on the marginal plots of the panel, which show that the $j$-sample's average \jbar\ value is higher than those of the two non-isolated samples, for similar baryonic mass distributions.

The change in slope in the \citetalias{ManceraPina2021} sample is likely due to the presence of dwarf galaxies in the sample. In fact, \citet{Kurapati2018} found that the angular momentum of dwarf galaxies is higher than what would be expected from the extrapolation of the \Mbar--\jbar\ relation for more massive galaxies. This suggests that the relation is a broken power law having a higher slope at the low-mass end of the relationship than at the higher-mass end. To investigate this, we derive the slope of the low-mass population of the \citetalias{Murugeshan2020} and \citetalias{ManceraPina2021} samples, and for the \citet{Kurapati2018} and \citet{Butler2017} samples of dwarf galaxies: the results of the regressions are outlined in \Cref{tb:fits}. As the table shows, the slopes of the low-mass populations are systematically higher than those of the high-mass populations.

To ensure that this is not an effect of the non-converging galaxies, we have restricted the analysis to only the converging galaxies of \citetalias{ManceraPina2021} sample for which convergence analysis is available; we identify 105 out of 151 galaxies meeting the convergence criterion set by the authors. Of the 46 non-converging galaxies, 80\% (i.e., 37 galaxies) are low-mass galaxies according to the mass criterion set above. We found slopes of $\alpha_{-} = 0.67\pm0.06$ and $\alpha_{+} = 0.55\pm0.06$, respectively for the low-mass and high-mass galaxies of the sample. Although these slopes agree within the errorbars, their consistency with the results of \Cref{tb:fits} argues in favour of the broken power law hypothesis.

\begin{table*}
    \centering
    \caption{The results of linear regressions to the different samples, separated by baryonic mass: the transition mass between low and high mass galaxies is $\rm M_{bar} = 10^{9.6}\,M_\odot$.}
    \label{tb:fits}
    \begin{tabular}{llccclccc}
    \hline
    \hline
         \multirow{2}[1]{*}{Sample} &  \multicolumn{4}{c}{Low baryonic mass} & \multicolumn{4}{c}{High baryonic mass} \\
    \cmidrule(lr){2-5} \cmidrule(lr){6-9}
           & size & $\alpha$ & $c$ & $\sigma$ & size & $\alpha$ & $c$ & $\sigma$ \\
    \hline
    AMIGA & \multicolumn{4}{c}{---} & 36 & $\alphaj$ & $\cj$ & $\sigmaj$\\
    M20 & \sizemurlow & $\alphamurlow$ & $\cmanlow$ & $\sigmamanlow$ & \sizemurhigh & $\alphamurhigh$ & $\cmurhigh$ & $\sigmamurhigh$\\
    MP21 & \sizemanlow & $\alphamanlow$ & $\cmanlow$ & $\sigmamanlow$ & \sizemanhigh & $\alphamanhigh$ & $\cmanhigh$ & $\sigmamanhigh$\\
    Kurapati+ 18 & \sizekur & $\alphakur$ & $\ckur$ & $\sigmakur$ & \multicolumn{4}{c}{---} \\
    Butler+ 17 & \sizebut & $\alphabut$ & $\cbut$ & $\sigmabut$ & \multicolumn{4}{c}{---} \\
    Kurapati+ 21 & \multicolumn{4}{c}{---} & \sizeuma & $\alphauma$ & $\cuma$ & $\sigmauma$\\
    \hline
    \end{tabular}
    \begin{tablenotes}
    \item * the initial sample was 14 but the outlier galaxy UGC 8508 was removed from the fit;
    \item ** the total size of the sample is 16 but a galaxy with a baryonic mass $10^{9.25}\,\rm M_\odot$ was removed from the fit.
    \end{tablenotes}
\end{table*}

\subsection{The Fall relation: gas vs. stellar discs}\label{sec:jvsm:gs}
\begin{figure*}
    \centering
    \includegraphics[width=\textwidth]{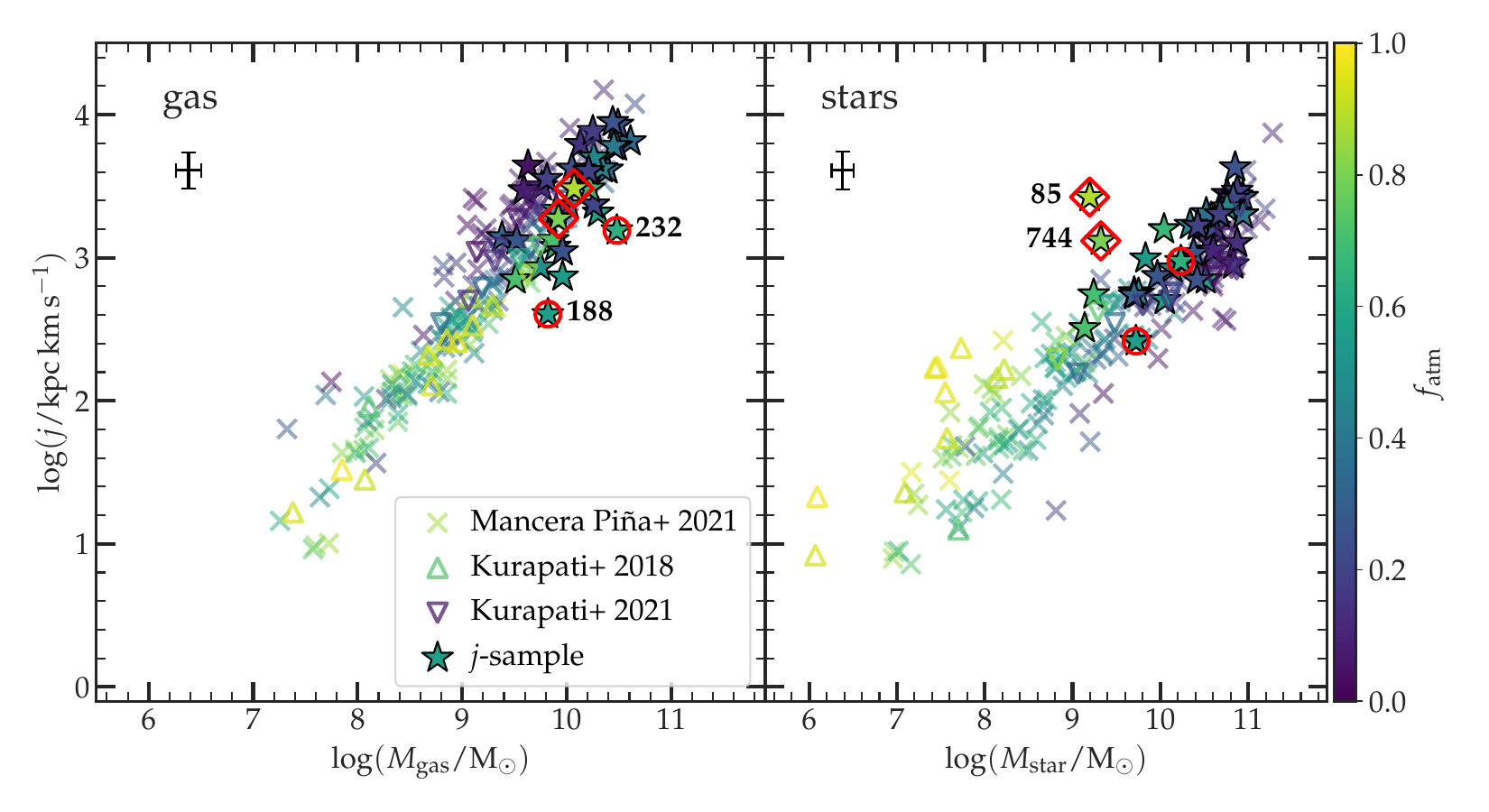}
    \vspace{-20pt}
    \caption{Specific angular momentum as a function of mass for the gas ({\it left}) and stellar ({\it right}) components, colour-coded with the atomic gas-to-total mass fraction. The red circles denote the galaxies with lower-than-average \jgas\ values, while those with higher-than-average \jstar\ values are marked with red diamonds. The numbers next to these markings correspond to the galaxies' CIG numbers. The typical error bars are shown at the top left corner of the panels.}
    \label{fig:jvsm-gs}
\end{figure*}

How is the angular momentum distributed among the galaxies' main components? In \Cref{fig:jvsm-gs} we present the specific angular momentum as a function of the mass, separately for the gas and stellar components and colour-coded by their gas fraction $f_{\rm atm}$ (fraction of atomic gas to total baryonic mass). For comparison, we overlay on the figure the samples of \citetalias{ManceraPina2021}, \citet{Kurapati2018} and \citet{Kurapati2021}. As expected, the galaxies in the AMIGA angular momentum subset sit on the high-mass end with respect to the non-isolated galaxies, both in terms of \hi\ and stars. Furthermore, in terms of gas fraction, two striking trends appear for all the galaxy samples, consistently with results from \citet{ManceraPina2021a}: at fixed gas mass, galaxies with high \jgas\ tend to have a low gas fraction; conversely, at fixed stellar mass, galaxies with high \jstar\ tend to have a high gas fraction. On the other hand, while the $j$ values of the gas component of the AMIGA galaxies seem to agree with those of the non-isolated samples, we note that their stellar component presents a different trend: the \jstar\ values of most AMIGA galaxies are among the highest at a given stellar mass.

By evaluating the deviations of the $j$-sample galaxies from the line of best fit, for each of the stellar and gas relations, and setting a maximum scatter of $2\sigma_i$ from each relation (where $\sigma_i$ is the standard deviation of the scatters around the line of best fit for component $i$), we identify two outliers in each of the panels: CIG 188 \& 232 for the \jgas\ distribution, and CIG 85 \& 744 for the \jstar\ distribution. We find that the two galaxies with abnormally low \jgas\ values (marked with red circles in the figure) present ``normal'' \jstar\ values with respect to the rest of the samples. Similarly, the galaxies with atypically high \jstar\ values (marked with red diamonds) exhibit ``normal'' \jgas\ values. To ensure that these galaxies are not outliers because of technical biases, we compared their angular resolutions (listed in \Cref{tb:sources}) to the rest of the sample and found that they are not particularly less resolved than the galaxies in the angular momentum plane.

\begin{figure}
    \centering
    \includegraphics[width=\columnwidth]{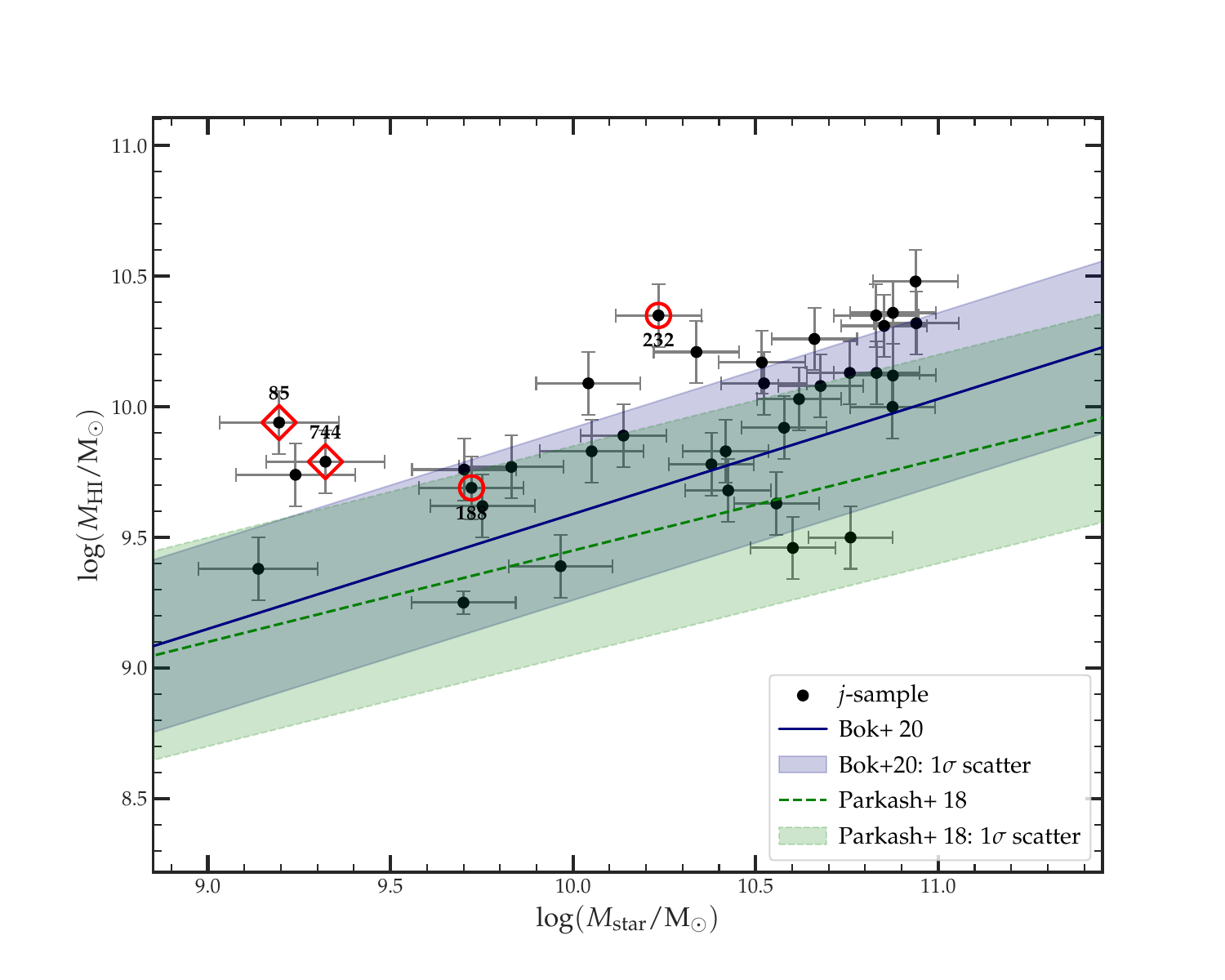}
    \vspace{-20pt}
    \caption{\hi\ vs. stellar mass of AMIGA galaxies. The blue and dashed green lines show the scaling relations respectively for 544 AMIGA galaxies \citep{Bok2020} and for a sample of $M_{\rm star}$-selected spiral galaxies \citep{Parkash2018}. The red circles and diamonds and the numbers are the same as in \Cref{fig:jvsm-gs}.}
    \label{fig:mhi-ms}
\end{figure}

Could the outlier galaxies either have an excess in their gas content (for CIG 188 \& 232) or a deficit in their stellar mass (for CIG 85 \& 744)?
To address the first part of the question, we compare the distribution of the \hi\ and stellar masses of the galaxies in the $j$-sample to trends found in \citet{Bok2020}, for a larger AMIGA sample, and in \citet{Parkash2018}, for a sample of spirals (\Cref{fig:mhi-ms}). The scaling relation of \citet{Bok2020} was derived by fitting a linear relation to 544 AMIGA galaxies of high-quality \hi\ profiles, selected from the Verley07b sample and whose stellar masses were estimated from mid-infrared {\it WISE} photometry. As for the \citet{Parkash2018}'s scaling relation, it was obtained from a sample of 600 optically-selected spiral galaxies of redshift $z\leq0.01$, with a completeness of 99\%. Similarly to \citet{Bok2020}'s sample galaxies, the stellar masses of the spirals in \citet{Parkash2018} were also measured from {\it WISE} bands photometry. The figure shows that most $j$-sample galaxies sit above both relations although, as expected, an important fraction falls in the region prescribed by the \citet{Bok2020} relation. We particularly note that CIG 188 presents an average gas mass, while CIG 232 shows a high gas content with respect to its stellar mass. Furthermore, an inspection of the rotation curves of these galaxies in \Cref{fig:rotcur} (and also in \Cref{fig:btf} discussed in the next section) shows that CIG 188 presents very low rotation velocities, with a maximum as low as ${\sim}40$ \kms. These suggest that the deviation of CIG 232 from the \jgas\ plane could be caused by an excess in its \hi\ mass, while that of CIG 188 is likely due to its slow rotation. Since the \hi\ masses presented here are measured from single-dish observations \citep[see][]{Jones2018}, this implies that CIG 232 contains a significant amount of low-density gas in its outer regions that is not seen in its kinematic maps. The  total mass \hi\ derived from the galaxy's integrated map reveals that $33.2\%$ of its single-dish flux is not recovered by the interferometric observations, higher than the median of $(19.0\pm5.1)\%$ for the entire $j$-sample. 
The missing gas could be in the form of faint \hi\ envelopes, similar to that around M83 \citep{Heald2016}; this is all the more possible since these galaxies are isolated, hence have fewer chances of seeing their envelopes disrupted.

Regarding CIG 85 \& 744, \Cref{fig:mhi-ms} shows that these galaxies have significantly higher gas masses for their low stellar masses -- they are, in fact, among the the galaxies with the lowest stellar masses. This makes them the highest gas fractions ($f_{\rm atm}>0.8$) in the $j$-sample. Two possibilities arise for these galaxies: their deviation from the \jstar-$M_\star$ relation is either caused by their low $M_\star$ (horizontal deviation), or by their high \jstar\ values (vertical deviation). CIG 85 and 744 are respectively classified as an irregular and a late-type spiral, with CIG 744 hosting an AGN in its centre \citep{Hernandez-Ibarra2013}. Furthermore, CIG 85 presents highly disturbed optical and \hi\ morphologies, leading \citet{Sengupta2012} to argue that the galaxy may have undergone minor mergers in the recent past. A vertical shift could be explained by the galaxies' high gas fractions, which confer them high \jstar\ values \citep[see, e.g.][]{Lutz2018,ManceraPina2021a}. The particularly low inclination of CIG 85 (${\sim}16\dg$) could also lead to an overestimation of its rotation velocity, pushing the galaxy upwards in the $j$ plane. On the other hand, one could be tempted to attribute the horizontal shift to the lower mass-to-light ratio adopted for these galaxies' types (see \Cref{sec:data:wise}). We note that independent measurements in the literature quote a maximum stellar mass of $2.4\times10^9\rm\,M_\odot$ \citep[from $L_{\rm B}$ photometry]{Sengupta2012} and $1.5\times10^9\rm\,M_\odot$ \citep[from SED fitting]{Chang2015} respectively for CIG 85 and 744, consistent with the values measured in this work. Furthermore, for these galaxies to fall on the relation at these \jstar\ values, their $M/L_{W1}$ values would have to be increased to respectively 4.5 and 1.7, which is much higher than allowed. This therefore discards the second possibility, allowing us to conclude that these two outliers possess a lot more stellar angular momentum for their optical size, possibly as a result of their high gas fractions.

\section{Discussion}\label{sec:discussion}
\begin{figure*}
    \centering
    \includegraphics[width=\textwidth]{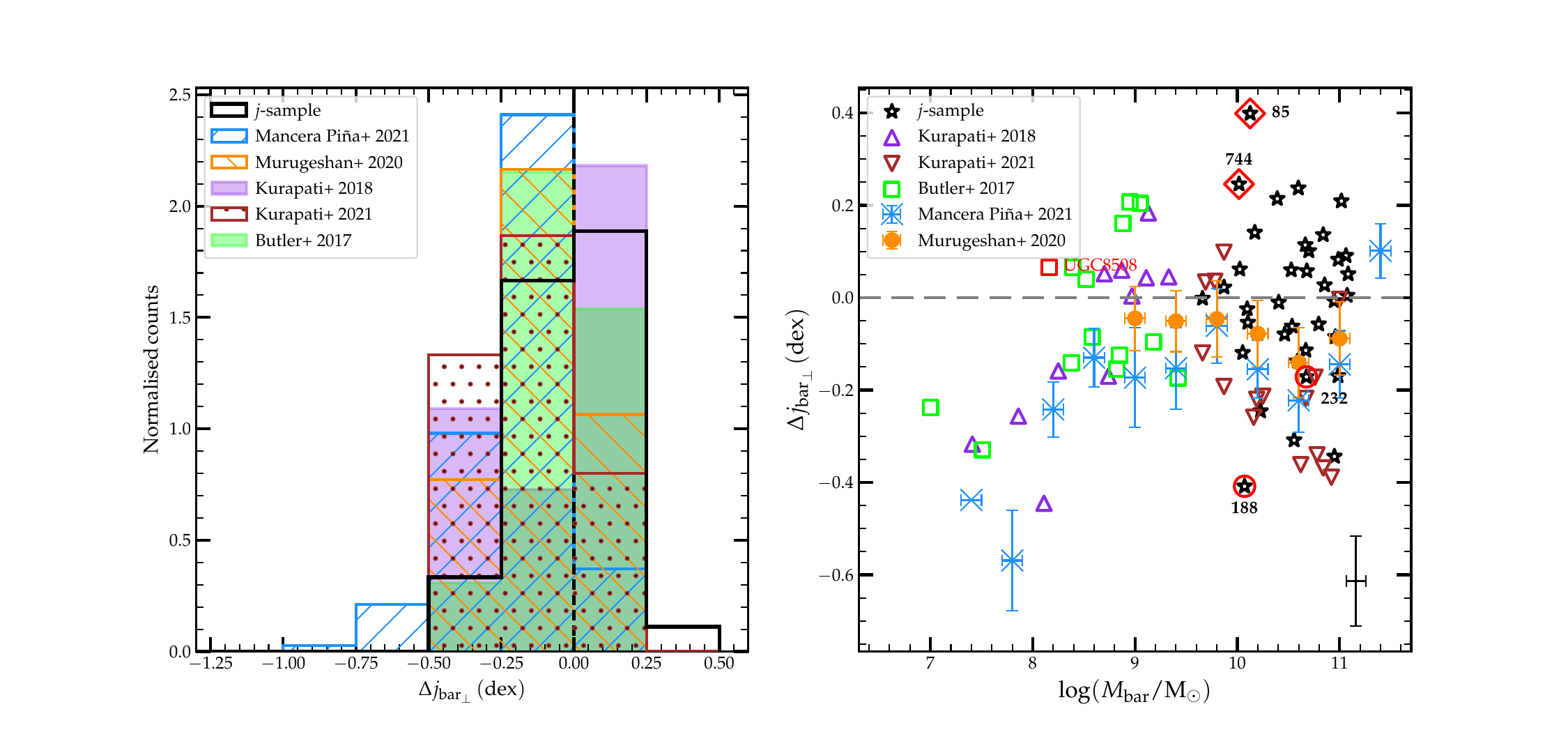}
    \caption{{\it Left}: distribution of the specific baryonic angular momentum's deviation $\Delta j_{{\rm bar}_\perp}$ of the different samples from the fit of the isolated $j$-sample.  {\it Right}: $\Delta j_{{\rm bar}_\perp}$ as a function of the baryonic mass. For clarity, the \citetalias{Murugeshan2020} and \citetalias{ManceraPina2021} samples were binned and only the median values and standard deviations of the bins are shown. The typical error bars are shown at the bottom right corner of the panel. The red circles and diamonds and the numbers are the same as in \Cref{fig:jvsm-gs}.}\label{fig:deltaj}
\end{figure*}

The AMIGA sample is, unlike field galaxies, a nurture-free sample in the sense that it is constituted of galaxies that have not undergone any major interaction in the past ${\sim}3$ Gyr \citep{Verdes-Montenegro2005a}. Therefore, it represents a reference for evaluating the effects of the environment on the angular momentum. Under this consideration, the results of \Cref{fig:jvsm-bar}, showing that the AMIGA galaxies possess higher $j$, further support the initial hypothesis: isolated galaxies contain higher angular momentum than their non-isolated counterparts, mainly because the effects of environmental processes on their kinematics are less important. In other words, they still possess a higher fraction of their initial angular momentum because they have undergone fewer major interactions than their counterparts in denser environments.
To further demonstrate this, we present in \Cref{fig:deltaj} the distribution of the orthogonal deviation\footnote{the non-vertical, perpendicular deviation of a given data point from the $j$-sample line of best fit, measured as the separation between the point and the best fit line.} of \jbar\ from the $j$-sample's scaling relation, for all the comparison samples. Most galaxies in the samples have \jbar\ such that $-0.5 \leq \Delta j_{{\rm bar}_\perp} \leq 0.0$, with only a low fraction (24\%) of galaxies having $\Delta j_{{\rm bar}_\perp} > 0$. The right panel of the figure shows the same deviation $\Delta j_{{\rm bar}_\perp}$, but plotted as a function of the baryonic mass. As expected from \Cref{fig:jvsm-bar}, the deviation is higher for low-mass galaxies (\citealt{Butler2017} and \citealt{Kurapati2018} samples), with a general trend of $\Delta j_{{\rm bar}_\perp}$ increasing with the baryonic mass from $-0.6$ dex to about 0.2 dex. 

Particularly, the median $\Delta j_{{\rm bar}_\perp}$ values of the \citetalias{Murugeshan2020} and \citetalias{ManceraPina2021} galaxies agree within their errorbars and are on average negative throughout the probed baryonic mass range. This further supports the hypothesis that the specific angular momentum of the isolated AMIGA galaxies is overall higher than those of the non-isolated galaxies. We also show the deviation of the isolated $j$-sample from the linear fit, exhibiting a significant scatter around the best fit line; obviously (and as expected), the deviation is among the largest for the outlier galaxies. To quantify the amplitude of the deviations, we present in \Cref{tb:deltaj} the standard deviation of the distributions of $\Delta j_{{\rm bar}_\perp}$ for each of the samples. The distribution of $\Delta j_{{\rm bar}_\perp}$ for the $j$-sample has a standard deviation (0.17 dex) larger than those of the \citetalias{ManceraPina2021} and \citet{Kurapati2018} samples, respectively. However, after removing the four outliers from the sample, the standard deviation of $\Delta j_{{\rm bar}_\perp}$ decreases to 0.14 dex, the lowest of all six samples. This further shows that the large scatter in the $j$-sample is caused by the four galaxies (11\% of the sample) exhibiting either peculiar rotational velocities or a high gas content. However, this low scatter remains mathematically consistent with what is expected from the removal of the outliers, and is not significantly lower than that of the comparison samples. This is inconsistent with previous studies conducted on the properties of the galaxies in the AMIGA sample, finding that their parameters sensitive to environmental processes tend to present more uniform values \citep{Lisenfeld2007,Lisenfeld2011,Espada2011,Sabater2012}. This suggests that the AMIGA galaxies could present a larger diversity in their kinematics than previously thought.

\begin{table}
\centering
\caption{The mean values and standard deviations in the distributions of $\Delta j_{{\rm bar}_\perp}$ of the different samples.}\label{tb:deltaj}
\begin{tabular}{l c c c}
\hline
\hline
Sample & Size & $\mu_{\Delta j_{{\rm bar}_\perp}}$ & $\sigma_{\Delta j_{{\rm bar}_\perp}}$ \\
[2pt]
\hline
AMIGA & 36 & 0.00 & 0.17 \\
AMIGA (no outliers) & 32 & 0.00 & 0.14 \\
MP21 & 156 & -0.20 & 0.18 \\
M20 & 109 & -0.10 & 0.15 \\
Kurapati+ 2018 & 11 & -0.09 & 0.19 \\
Kurapati+ 2021 & 15 & -0.18 & 0.15 \\
Butler+ 2017 & 13 & -0.05 & 0.17 \\
\hline
\end{tabular}
\end{table}

\begin{figure}
    \centering
    \includegraphics[width=\columnwidth]{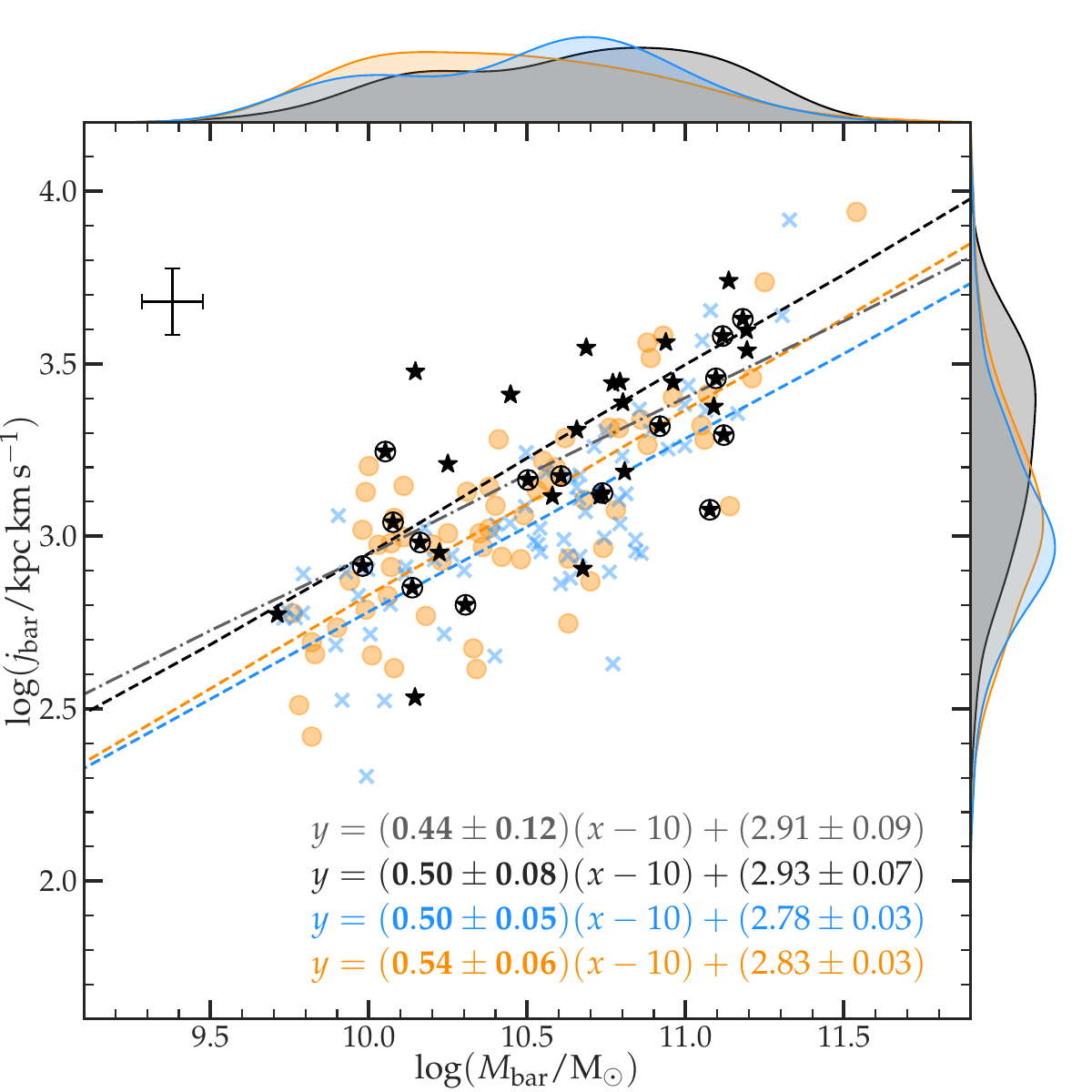}
    \caption{The right panel of \Cref{fig:jvsm-bar} reproduced with $M/L_{W1}=0.5$ for the $j$-sample galaxies. All symbols are the same as in \Cref{fig:jvsm-bar}.} \label{fig:jbar-ml}
\end{figure}

The stellar masses of the $j$-sample galaxies were derived from their {\it W1} magnitudes, with a mass-to-light ration of $M/L_{W1} = 0.35\pm0.05$. As noted in \Cref{sec:data:wise}, this is 30\% lower than the values adopted in \citetalias{ManceraPina2021}, the largest comparison sample used in this work. To ensure that the observed higher \jbar\ values are not solely due to the difference in the $M/L$ values, we perform a test by adopting the same $M/L$ as in \citetalias{ManceraPina2021}. In \Cref{fig:jbar-ml} we show the resulting \Mbar-\jbar\ relation, where only the stellar masses of the $j$-sample galaxies were altered. The change in the $M/L_{W1}$ value leads to an average increase of ($0.11\pm0.01$) dex in the baryonic masses of the galaxies, making the new line of best fit shift downward by ($0.02\pm0.03$) dex on average. However, the previously observed trend remains: \jbar\ is higher for the isolated galaxies than their non-isolated counterparts.

\subsection{AMIGA galaxies in the baryonic Tully-Fisher plane}\label{sec:disc:btf}
A simpler way to evaluate the {\it normality} of the rotation of disc galaxies is through the empirical baryonic Tully-Fisher relation (BTF), that links a galaxy's rotation velocity to its baryonic mass. Originally established between optical luminosity and the velocity width \citep{Tully1977}, the BTF was later translated into a tight linear relation between the flat part ($V_{\rm flat}$) of a disc rotation curve and its total baryonic mass over several mass orders of magnitude \citep{McGaugh2000,Verheijen2001a,Bell2001,McGaugh2015}. The BTF relation has widely been investigated and constrained in the literature, with authors often describing the rotation of galaxies by either the \hi\ line width or $V_{\rm flat}$ \citep[e.g.,][]{Noordermeer2007,Trachternach2009,Gurovich2010,McGaugh2012,Lelli2016a}.

In order to test whether the galaxies in the $j$-sample exhibit different rotation patterns than their non-isolated counterparts, we investigate their positions in the BTF plane. If they possess peculiar rotation velocities, they should stand out in the BTF plane. In other words, if their observed high \jbar\ values stem from overestimated rotation velocities, they should deviate from the BTF relation.

\begin{figure}
    \centering
    \includegraphics[width=\columnwidth]{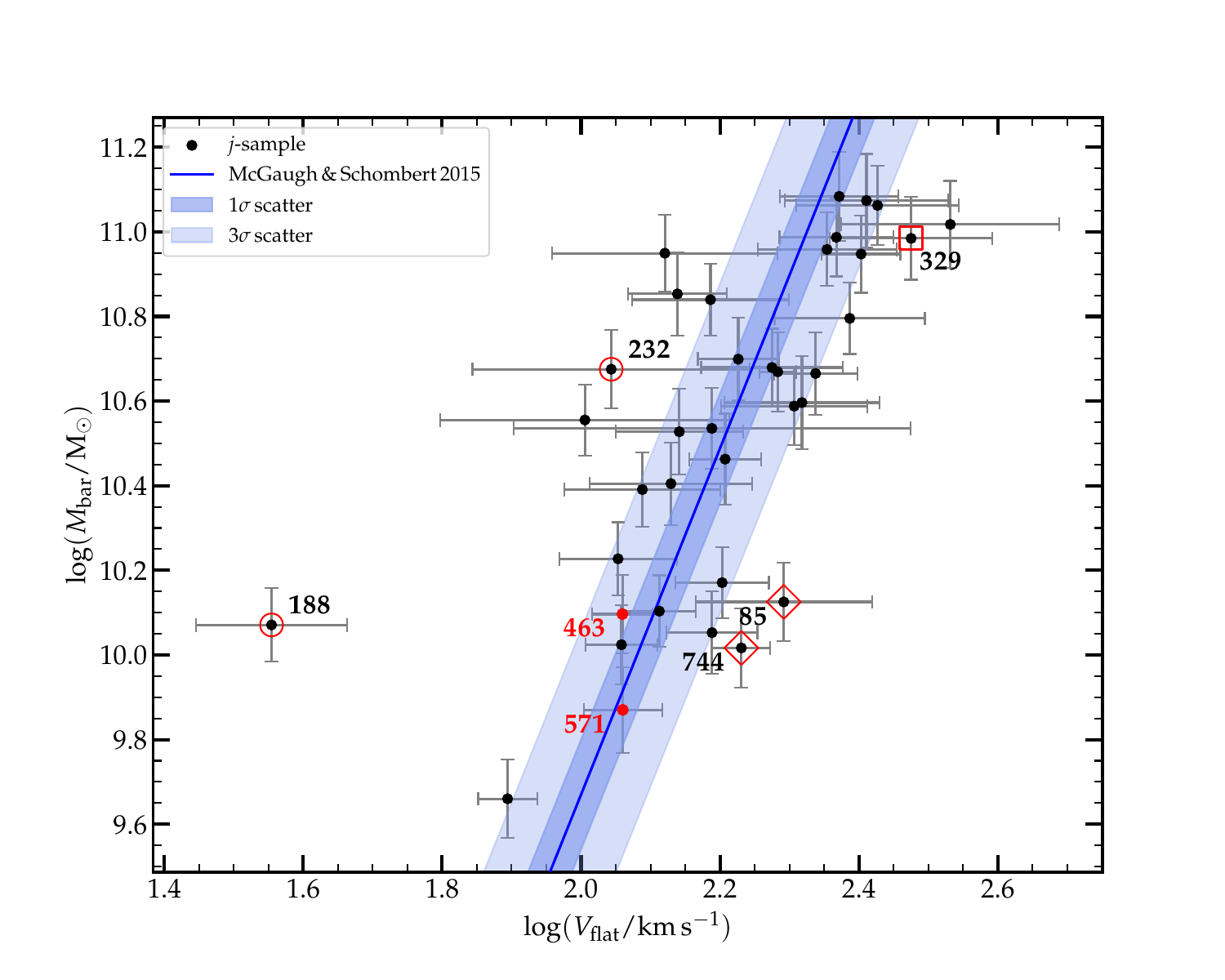}
    \vspace{-20pt}
    \caption{Baryonic Tully-Fisher relation of the $j$-sample compared with the fit from \citet{McGaugh2015}. The two galaxies with non-flat rotation curves are marked with red dots. The red circles and diamonds and the numbers are the same as in \Cref{fig:jvsm-gs}. CIG 329 (marked with a red square) is discussed in \Cref{sec:disc:btf}.}\label{fig:btf}
\end{figure}

From the rotation curves of the galaxies, we determine $V_{\rm flat}$ using the algorithm adopted in \citet{Lelli2016a} and described in \Cref{app:rotcur}. All galaxies in the sample reach the flat part of their rotation curve, except CIG 463 \& 571 who seem to be still rising. Nonetheless, \Cref{fig:btf} shows that these two galaxies lie within one standard deviation of the BTF relation derived by \citet{McGaugh2015}. It is, however, worth mentioning that an inspection of CIG 571's PV diagram (\Cref{app:velfields}) indicates a possible overestimation of its rotation velocities in its outer regions, possibly worth investigating further with deeper data. On the other hand, CIG 329 exhibits peculiar rotation velocities in the external regions, with the outermost part of its curve hinting increasing velocities. This is likely caused by the complex kinematics of the galaxy. In fact, its \hi\ maps and PV diagram reveal a severe warp in both sides of the galaxy's \hi\ disc, described by \citet{Spekkens2006} as symmetric and extreme. Although the best kinematic model for CIG 329's disc was yielded by a constant inclination, we do not discard the possibility that, in reality, the observed warps induce variations in the inclination of the \hi\ disc. 

As shown in \Cref{fig:btf}, the $j$-sample galaxies do not particularly favour high rotation velocities; instead, most sample galaxies are roughly evenly spread across the BTF relation, occupying both sides of the relation.

\subsection{Comparison with the CDM model}\label{sec:disc:cdm}
The $\Lambda$CDM cosmology predicts that the baryonic specific angular momentum can be written \citep[][assuming $H_0 = 70\,\rm km\,s^{-1}\,Mpc^{-1}$]{Obreschkow2014}:
\begin{equation}\label{eq:cdm}
    j_{\rm bar}/(10^3\, {\rm kpc\,km\,s^{-1}}) = k_{\rm f}\,[M_{\rm bar}/(10^{10} M_\odot)]^{2/3}
\end{equation}
where the coefficient $k_{\rm f} = 1.96\,\lambda\,f_j\,f_M^{-2/3}$ is function of the halo spin parameter $\lambda$, the baryon-to-halo specific angular momentum fraction $f_j$ and the baryon-to-halo mass fraction $f_M$. \citet{Obreschkow2014} made different considerations to approximate the values of the parameters; namely, the authors adopt $\lambda \approx 0.04\pm0.02$ (independent of the halo mass) from N-body simulations \citep{Maccio2008}, $f_j \approx 1.0\pm0.5$ based on simulations of Milky Way-like galaxies \citep{Stewart2013} and lastly $f_M \approx 0.05$. These values constrain the coefficient $k_{\rm f}$ to vary between 0.14 and 1.3, allowing to visualize the shape of \Mbar--\jbar\ relation as predicted by the model.

\begin{figure*}
    \centering
    \includegraphics[width=\textwidth]{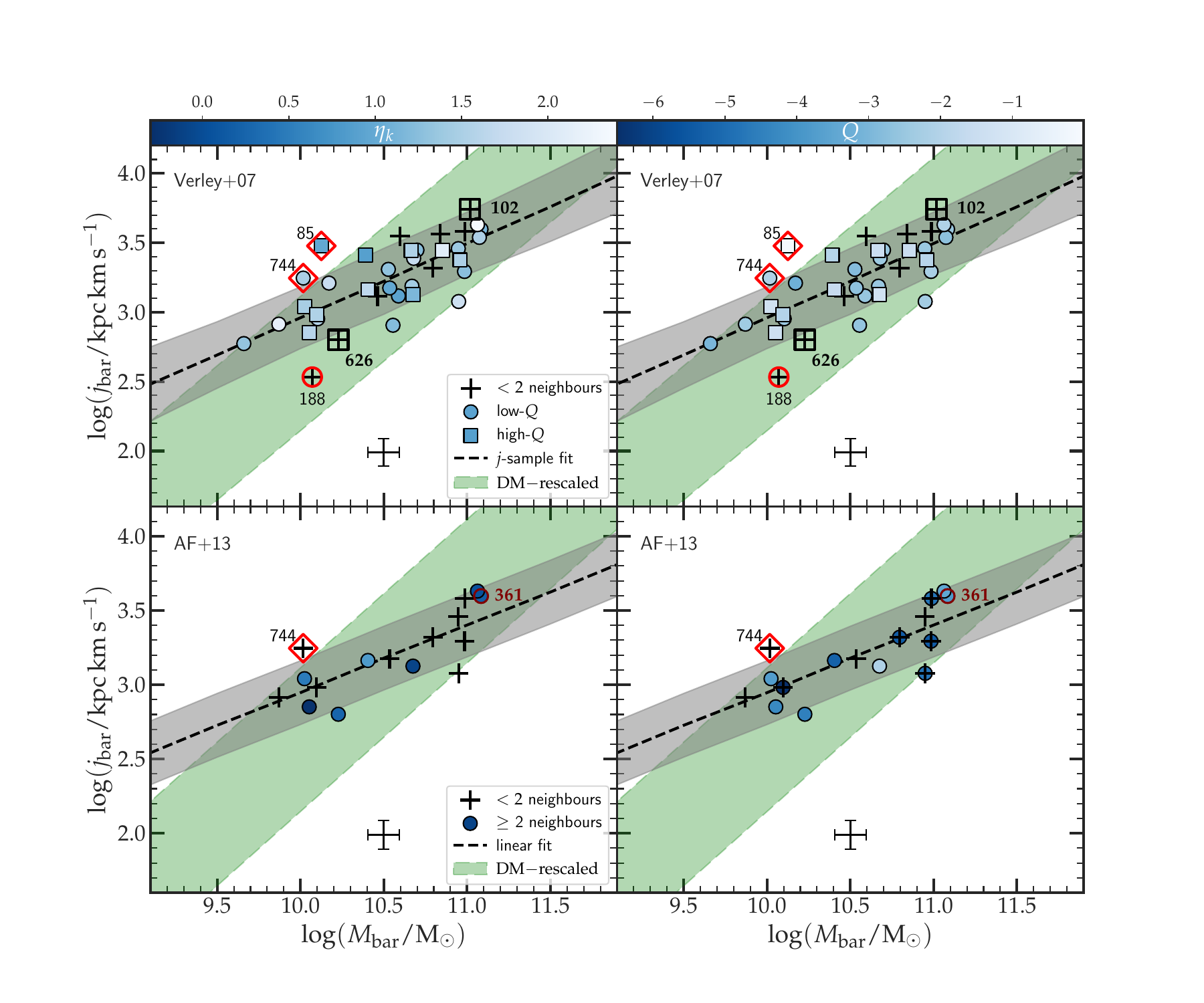}
    \vspace{-20pt}
    \caption{Comparison of isolated $j$-sample with the DM-rescaled model as a function of the isolation parameters: the local number density $\eta_k$ (left) and the total force $Q$ exerted by the galaxy's neighbours (right). The {\it top panels} use the isolation definition of \citet{Verley2007a} to distinguish between isolated and non isolated galaxies, while the {\it bottom panels} use that of \citet{Argudo-Fernandez2013} and only include low $Q$ galaxies. The only non-isolated galaxy contained in the \citet{Argudo-Fernandez2013}'s sample (CIG 361) is denoted by a dark red circle. The grey area denotes the standard deviation of the fit, and the green area the DM-rescaled model, whose width is determined by the $k_{\rm f}$ value (see \Cref{sec:disc:cdm}). The typical error bars are shown in the bottom of each panel. The red circles and diamonds and the numbers are the same as in \Cref{fig:jvsm-gs}.}\label{fig:jcdm}
\end{figure*}

We show in \Cref{fig:jcdm} a comparison of the AMIGA angular momentum sample's \Mbar--\jbar\ relation to the DM-rescaled model. The width of the model is determined by the value of the factor $k_{\rm f}$ of \Cref{eq:cdm} which, as noted above, varies between 0.14 and 1.3. Most AMIGA galaxies lie within the range predicted by the model, with the exception of two galaxies at the lower mass end (CIG 85 \& 744), previously found to have higher-than-average \jstar\ values (see \Cref{sec:jvsm:gs}). Furthermore, as noted in \Cref{sec:jvsm:isol}, the slope of the AMIGA \Mbar--\jbar\ relation is shallower than the theoretical prediction of $\alpha \sim 2/3$ by about 22\%. It is worth mentioning that the model neglects the dependency of $f_j$ and $f_M$ with the halo mass, and therefore gives a theoretical prediction independent of the actual baryonic mass.

\subsection{On the relation between $j$ and galaxy isolation}\label{sec:disc:dep}
One of the major results of this work is that the angular momentum of isolated galaxies, even those of the high-$Q$ sample (\Cref{sec:data:amiga}) considered as not strictly isolated, is on average higher than that of their non-isolated counterparts. This hints that the environment might play an important role in removing the angular momentum of galaxies through interactions. However, a question that remains unanswered is whether the position of a galaxy in the angular momentum space is correlated with its degree of isolation. In other words, do more isolated galaxies have lower angular momentum than their less isolated counterparts? To investigate this, we separate the isolated $j$-sample into three subgroups: in the first, we consider the galaxies having less than two neighbours in the \citet{Verley2007b} catalogue and, as a consequence, have undetermined $\eta_k$ values. We identify seven galaxies in this category, including CIG 188, which was previously found to have abnormally low \jgas\ values. The second and third bins comprise respectively the low-$Q$ and high-$Q$ subsamples defined in \Cref{sec:data:amiga}. We colour-code these with the said isolation parameters in each of the two panels of \Cref{fig:jcdm}. With the exception of CIG 102 and 626 (labelled with black empty squares in the figure), the high-$Q$ sample is by definition less isolated than the low-$Q$ sample, with the first bin (of zero or one neighbour) containing the most isolated galaxies. These two exceptions are classified as less isolated because, although they have less than two neighbours, their tidal force parameter is $Q>-2$.
In \Cref{fig:jcdm}, the distribution of galaxies of the different subgroups in the parameter space shows no correlation between the angular momentum and either of the isolation parameters $\eta_k$ and $Q$. This is translated by the absence of any clear trends observed between the three subgroups. In particular, the most isolated subgroups (the low-$Q$ and few neighbours subgroups) do not appear to have the highest angular momentum of the sample, nor do they present a distinct trend in the parameter space; instead, they are ``randomly'' distributed along the \jbar$-$\Mbar\ line of best fit.
Moreover, for a more complete analysis, we consider the more robust, three-dimensional definition of the isolation parameters in \citet{Argudo-Fernandez2013}. In some cases, the isolation parameters derived by \citet{Argudo-Fernandez2013} significantly differ from those of \citet{Verley2007b} due to both the differences in the search for neighbours and the evaluation of exerted force parameter $Q$. The 16 galaxies in the $j$-sample whose isolation parameters were evaluated by the authors are plotted in the lower panel of the \Cref{fig:jcdm}. Similarly to the top panels, no clear trend is found in the distribution of the isolation parameters along the $j$ plane.

These results, combined with the fact that isolated galaxies possess a higher angular momentum with respect to their non-isolated counterparts, suggest that there is a threshold density beyond which the effects of interactions become important in removing the angular momentum in galaxies. This implies that minor interactions between a galaxy and its neighbours will not considerably remove its angular momentum, unless the tidal forces that it experiences are important enough. Likewise, if the galaxy resides in a low-density environment, the effects of the said environment on its angular momentum will not be significant. However, given the reduced size of the sample used in the present work and the modest robustness of the isolation parameters (see discussion in the appendix of \citealt{Jones2020}), further investigation is necessary to confirm the existence of such threshold density. New and upcoming surveys with the MeerKAT \citep{Jonas2016} and ASKAP telescopes \citep[e.g., the {\sc Wallaby} survey;][]{Koribalski2020} will offer the possibility to investigate this by targeting galaxies in a wide range of environments.

\subsection{The disc stability of AMIGA galaxies}\label{sec:stability}
\begin{figure}
    \centering
    \includegraphics[width=\columnwidth]{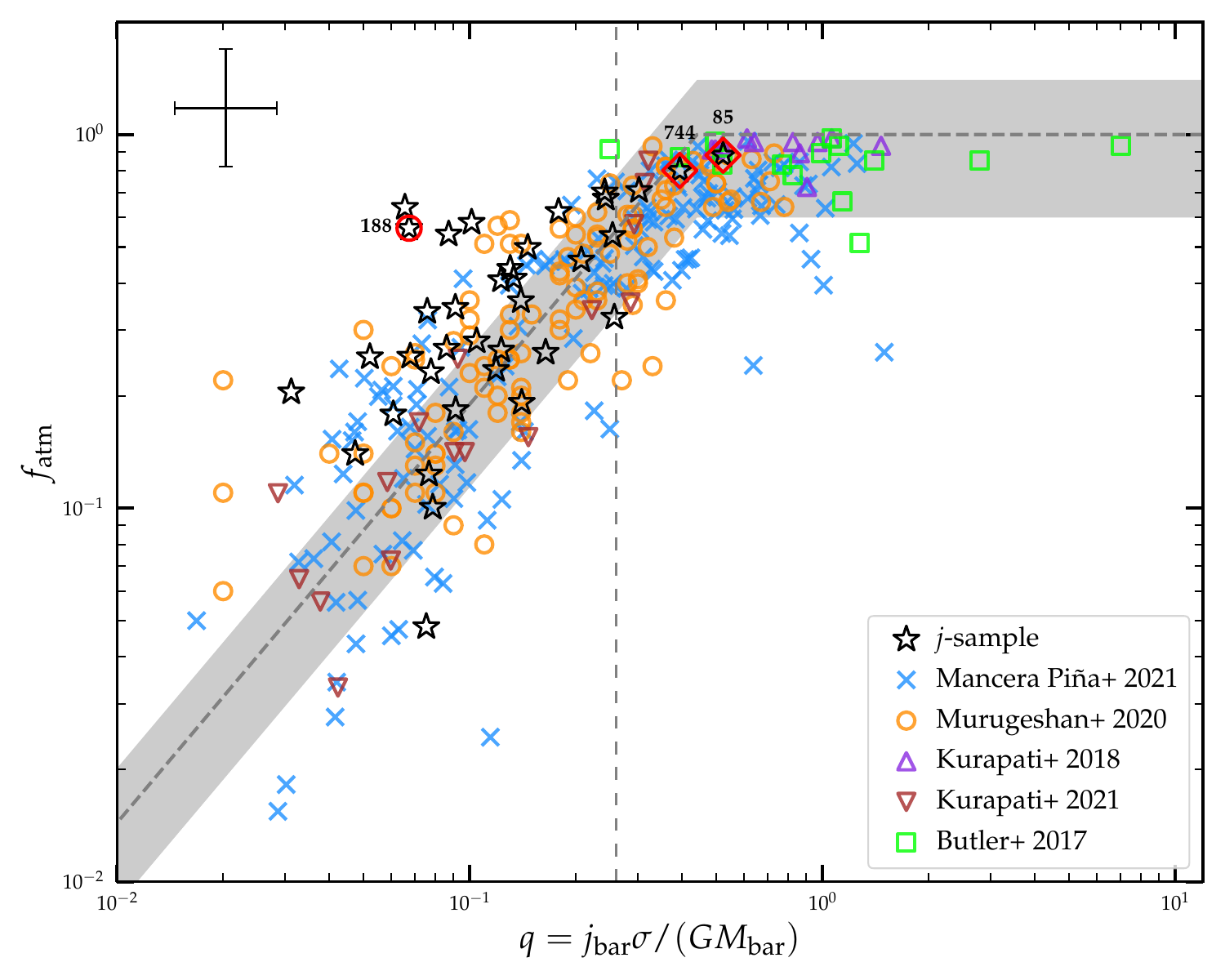}
    \vspace{-20pt}
    \caption{The atomic gas fraction versus the global disc stability parameter. The grey dashed line represents the model from \citet{Obreschkow2016}, and the shaded region its 40\% scatter. The vertical dashed line corresponds to $q = (\sqrt{2}\,e)^{-1}$, which theoretically separates the unstable (left) and the stable (right) discs. The typical error bars are shown at the top left corner of the figure. The red circles and diamonds and the numbers are the same as in \Cref{fig:jvsm-gs}.}\label{fig:fvsq}
\end{figure}

The stability of galaxy discs is an important parameter in their ability to form stars. In fact, both numerical simulations and observations argue that unstable galaxy discs are more susceptible to host higher star formation rates than their more stable counterparts \citep[e.g.,][]{Martin2001,Dutton2012,Stevens2016}, since star formation is thought to be provoked by the collapse of the neutral gas which is converted into stars via molecular gas. A widely accepted method of quantifying the stability of the galaxy disc is through the so-called Toomre criterion \citep{Toomre1964} for an axisymmetric rotating disc, which predicts that a disc is locally stable only if the pressure gradient at small scales is large enough to overcome the large scale centrifugal forces. For a galaxy disc of neutral atomic gas, the criterion is translated by the Toomre parameter
\begin{equation}
    Q_{\rm atm} = \frac{\kappa\,\sigma_{\rm atm}}{\pi G \Sigma_{\rm atm}},
\end{equation}
where $\kappa$ is the local epicyclic frequency, $G$ the gravitational constant, $\sigma_{\rm atm}$ and $\Sigma_{\rm atm}$ are respectively the local radial velocity dispersion and local surface density of the atomic gas. A stable, poorly star-forming galaxy disc is such that $Q_{\rm atm} > 1$, whereas $Q_{\rm atm} < 1$ corresponds to a more efficient star-forming, unstable disc. Building on this, \citet{Obreschkow2014} introduced a dimensionless {\it global} disc stability parameter, function of the specific angular momentum, the velocity dispersion and the mass of the disc:
\begin{equation}
    q = \frac{j_{\rm bar}\,\sigma_{\rm atm}}{GM_{\rm bar}}.
\end{equation}
Later, \citet{Obreschkow2016} established that the atomic gas fraction $f_{\rm atm}$ varies with the global parameter $q$, such that $f_{\rm atm} = \min{\{1, 2.5q^{1.12}\}}$. Interestingly, they also found that galaxies from various samples and including different morphologies tend to follow the model-based predictions of the $f_{\rm atm} {=} f(q)$ relation. We overlay in \Cref{fig:fvsq} the AMIGA galaxies as well as the non-isolated samples on the \citet{Obreschkow2016}'s model. Based on the discussion therein and for consistency in the comparisons, we adopt $\sigma_{\rm atm} {=} 10\,\rm km\,s^{-1}$. We also show in the figure the value $q = (\sqrt{2}\,e)^{-1}$, which \citet{Obreschkow2016} worked out to correspond to $Q_{\rm atm}\approx 1$, that is, the theoretical value at which galaxy discs turn from unstable to stable. 

Although the AMIGA galaxies seem to follow the trend of the theoretical model, it is interesting to note that more than half of them (19 out of 36) have atomic gas fractions higher than what the model predicts, with $f_{\rm atm}$ values beyond the 40\% margin allowed by the model. This is intriguing since it suggests that these galaxies have larger reservoirs of \hi\ than their angular momentum allows. In other words, their stability parameter $q$ is lower for their gas content, locating them on the left-hand side of the stability line where their gaseous discs are predicted to be unstable. Viewed from this angle, a large fraction of the galaxies in the isolated $j$-sample (especially those at lower $q$ values) can be interpreted as discs susceptible of collapsing on short timescales to form stars \citep{Obreschkow2016}. In light of all the assumptions made above, it is likely that the model is not perfectly suited for the highly isolated galaxies like those in the AMIGA sample, although \citet{Obreschkow2016} found it to well describe moderately isolated galaxies such as those of the THINGS and HIPASS \citep[\hi\ Parkes All-Sky Survey;][]{Meyer2004} samples. In either case, the high gas content of the AMIGA galaxies for such moderate $q$ values forces us to consider the possibility that many of these galaxies are (or have been) accreting an important amount of \hi\ in a recent period of their evolutionary phase. This is further supported by a comparison of their gas fraction with the other samples: they exhibit higher $f_{\rm atm}$ values compared to all other five samples, consistently with the results of \Cref{fig:mvsmfrac-lit} and those found in \citet{Jones2018}. Given the high level of isolation of AMIGA galaxies, such accretion would most likely happen through gas infall from the intergalactic medium, as opposed to accretion through galaxy mergers. Currently, the best direct method to obtain evidence of accretion is through high sensitivity mapping of these galaxies in search for companion \hi\ clouds, extra-planar gas or extended warps \citep{Sancisi2008}. High sensitivity \hi\ data combined to existing multi-wavelength data will allow us to further investigate this in the future.

The two galaxies discussed in \Cref{sec:jvsm:gs} (CIG 85 \& 744) to have low stellar masses with respect to their angular momentum appear to be among the few ``stable'' discs in the isolated $j$-sample, located near the region populated by mainly dwarf galaxies. However, they are not the least massive galaxies of the sample, and neither are they dwarf. We argue that their location in the parameter space is simply a direct consequence of their position in the \jbar$-$\Mbar\ relation: they exhibit a high baryonic angular momentum for an intermediate baryonic mass. As for the outliers exhibiting lower \jgas\ (red circles in the figure), they are the most discrepant galaxies with respect to the \citet{Obreschkow2016} model. As discussed above, these galaxies could be candidates for galaxies which have recently experienced gas accretion.

\section{Summary}\label{sec:summary}
We have investigated the behaviour of the angular momentum for isolated galaxies through the $j$-mass relation, using 36 galaxies drawn from the AMIGA sample. The aim of this study was to highlight the effects of the environment on the amount of baryonic angular momentum in galaxies, particularly testing whether interactions can remove galaxies' angular momentum as expected from our current understanding of galaxy evolution. In other words, we aimed to investigate whether isolated galaxies retain a higher fraction of their angular momentum, which would translate to these galaxies having higher $j$ values than their non-isolated counterparts. The main results of this work are as follows:
\begin{enumerate}
    \item At a fixed baryonic mass, the isolated galaxies of the AMIGA sample possess a higher specific angular momentum than their non-isolated counterparts (see \Cref{fig:jvsm-bar}). This constitutes direct evidence of the role of the environment in removing angular momentum from galaxies, predicted by numerical simulations \citep[e.g.,][]{Hernquist1995,Lagos2017}. In fact, galaxies in the AMIGA sample have, in theory, not undergone any major galaxy-galaxy interactions during the last ${\sim}3$ Gyr \citep{Verdes-Montenegro2005a}, reducing their loss of angular momentum with respect to interacting galaxies.
    \item High baryonic mass galaxies (${\gtrsim}10^9\,M_\odot$) are best fitted with a shallower power law compared to their lower mass counterparts. Consequently, lower mass galaxies (${\lesssim}10^9\,M_\odot$) exhibit a steeper power law; this change of slope is consistent with the broken power-law relation found by previous studies \citep{Butler2017,Kurapati2018}, where the angular momenta of low-mass galaxies deviate from the extension of the $j$-mass relation of more massive spirals.
    \item For the atomic gas component of all galaxies considered in this study (isolated and non-isolated), the specific angular momentum of gas-rich galaxies decreases with increasing gas fraction, for a fixed gas mass (\Cref{fig:jvsm-gs}). The reverse trend is seen for the stellar component, with the specific angular momentum increasing with the gas fraction at a given stellar mass. This is a consequence of the \jbar$-$\Mbar\ relation: at a fixed gas mass, gas-poor galaxies are more massive (in terms of baryons) than their gas-rich counterparts whereas, at a fixed stellar mass, it is the opposite.
    \item Most AMIGA galaxies included in this study agree with the DM-rescaled model in the \jbar$-$\Mbar\ plane, although their power-law slope ($\alphaj$) is ${\sim}30\%$ lower than the predicted slope (\Cref{fig:jcdm}). However, to effectively test whether isolated galaxies agree in general with the DM-rescaled model requires not only a broader range of baryonic mass, but also a tighter constraint on the width of the $f_{\rm atm} = f(q)$ model of \citet{Obreschkow2016}. We also find that all strictly isolated galaxies (i.e, galaxies with no identified neighbour in the optical) lie within the range predicted by the DM-rescaled model. However, no clear correlation was found between the position of the AMIGA galaxies on the \jbar$-$\Mbar\ relation and either of the $\eta_k$ and $Q$ isolation parameters.
    \item Four isolated galaxies were found to exhibit abnormal amounts of stellar or gaseous angular momentum (\Cref{fig:jvsm-gs}). The analysis of the kinematics and gas content of these galaxies shows that three possess high gas contents, while the last presents significantly low rotation velocities.
\end{enumerate}

These results, particularly the discrepancy between the AMIGA and non-isolated samples in the \jbar$-$\Mbar\ plane (see \Cref{fig:jvsm-bar}), provide clear evidence of the role of the local environment in removing angular momentum from galaxies, as suggested by previous studies \citep[e.g.,][]{Lagos2017}. However, one limitation of the present study is the lack of investigation of individual environmental processes that might affect the total angular momentum of the sample galaxies. For example, processes such as galactic winds and cold mode accretion are predicted to increase angular momentum \citep[e.g.,][]{Brook2012,Danovich2015}. Accounting for these individual processes, as well as targeting isolated galaxies of lower baryonic masses are interesting avenues for future studies. Furthermore, one consideration made in this study consisted of approximating the circular velocities of the stars to those of the gas. Although this approximation is appropriate for the large baryonic masses of the studied galaxies, the discrepancy seen in some outliers could be resolved by independently measuring their stellar velocities from spectroscopic IFU observations.


\section*{Acknowledgement}
We thank the anonymous reviewer for their valuable  suggestions and constructive feedback on the manuscript, which helped to improve the quality and clarity of the paper. This work used the Spanish Prototype of an SRC \citep[SPSRC;][]{Garrido2021} service and support funded by the Spanish Ministry of Science and Innovation (MCIN), by the Regional Government of Andalusia and by the European Regional Development Fund (ERDF).
AS, LVM, KMH, JG and SS acknowledge financial support from the grants SEV-2017-0709 funded by MCIN/AEI/10.13039/501100011033.
AS, LVM, JG and SS received further support from the grants RTI2018-096228-B-C31 and PID2021-123930OB-C21 funded by MCIN/AEI/10.13039/501100011033, by ``ERDF A way of making Europe" and by the European Union.
Lastly, part of the work of LVM, JG and SS was funded by the IAA4SKA grant (Ref. R18-RT-3082) from the Economic Transformation, Industry, Knowledge and Universities Council of the Regional Government of Andalusia and the European Regional Development Fund from the European Union. 

\section*{Data Availability}
The data underlying this article are available in  \Cref{tb:angmom} of this article and in the online supplementary material. The datacubes and kinematic products are available on request.



\bibliographystyle{mnras}
\bibliography{references} 


\appendix

\section{Posterior distribution of the fit parameters}\label{app:posterior}
To determine the best-fit parameters for the $j$-sample and other small size samples in this study, we have made use of the Student-$t$ distribution of probability density function
\begin{equation}\label{eq:st}
    p(y|\mu,\sigma,\nu) = \frac{\Gamma(\frac{\nu+1}{2})}{\sqrt{\nu\pi}\,\Gamma(\frac{\nu}{2})}\frac{1}{\sigma}\left[1 + \frac{\left(\frac{y-\mu}{\sigma}\right)^2}{\nu}\right]^{-\frac{\nu+1}{2}},
\end{equation}
where $\mu$, $\sigma$ and $\nu$ respectively represent the mean, standard deviation and degrees of freedom; the Gamma function is written as
\begin{equation}
    \Gamma(x) = \int_0^{\infty}{t^{x-1}e^{-t}dt} = (x-1)\Gamma(x-1).
\end{equation}
As noted in \citet{Shah2014}, the Student-$t$ distribution is a general, more flexible form of the Gaussian distribution, with the additional parameter $\nu$. Besides maintaining the advantages of Gaussian distributions, the Student-$t$ processes were shown to provide more robust results when accounting for outliers \citep[e.g.,][]{Shah2014,Tracey2018}.

In practice, the fitting method is as follows:
\begin{itemize}
    \item[-] the regression coefficients $\alpha$ and $c$ of \Cref{eq:j_eq} were given Gaussian priors of standard deviation 4 and centres 1 and 2 respectively: i.e, $\alpha \sim \mathcal{N}(1,4)$ and $c \sim \mathcal{N}(2,4)$;
    \item[-] the distribution of the {\it vertical} intrinsic scatter $\sigma$ was modelled by an exponential prior of coefficient 1: $\sigma \sim \rm Exp(1)$;
    \item[-] we chose a half-normal distribution of standard deviation 5 for the degrees of freedom: $\nu \sim \mathcal{H}(5)$. This parameter essentially sets the extent of the distribution’s tails, with $\nu=1$ corresponding to the heaviest tails while $\nu\rightarrow\infty$ converges to a normal distribution. By choosing a half-normal distribution, we aim to constrain tails of the likelihood's distributions to be heavier than a normal distribution, hence accounting for the outliers in the data;
    \item[-] next, the likelihood of the $\log{j_{\rm bar}}$ values is explored with a Student-$t$ distribution as defined in \Cref{eq:st}, with a mean $\mu \sim \alpha (\log{M_{\rm bar}} - 10) + c$, a standard deviation $\sigma$ and degree of freedom $\nu$;
    \item[-] finally, 4000 Markov chains are randomly drawn to determine the posterior, from which the best fit values of the regression parameters are derived.
\end{itemize}

\begin{figure}
    \centering
    \begin{minipage}[c]{\columnwidth}
     \centering
     \includegraphics[width=\columnwidth]{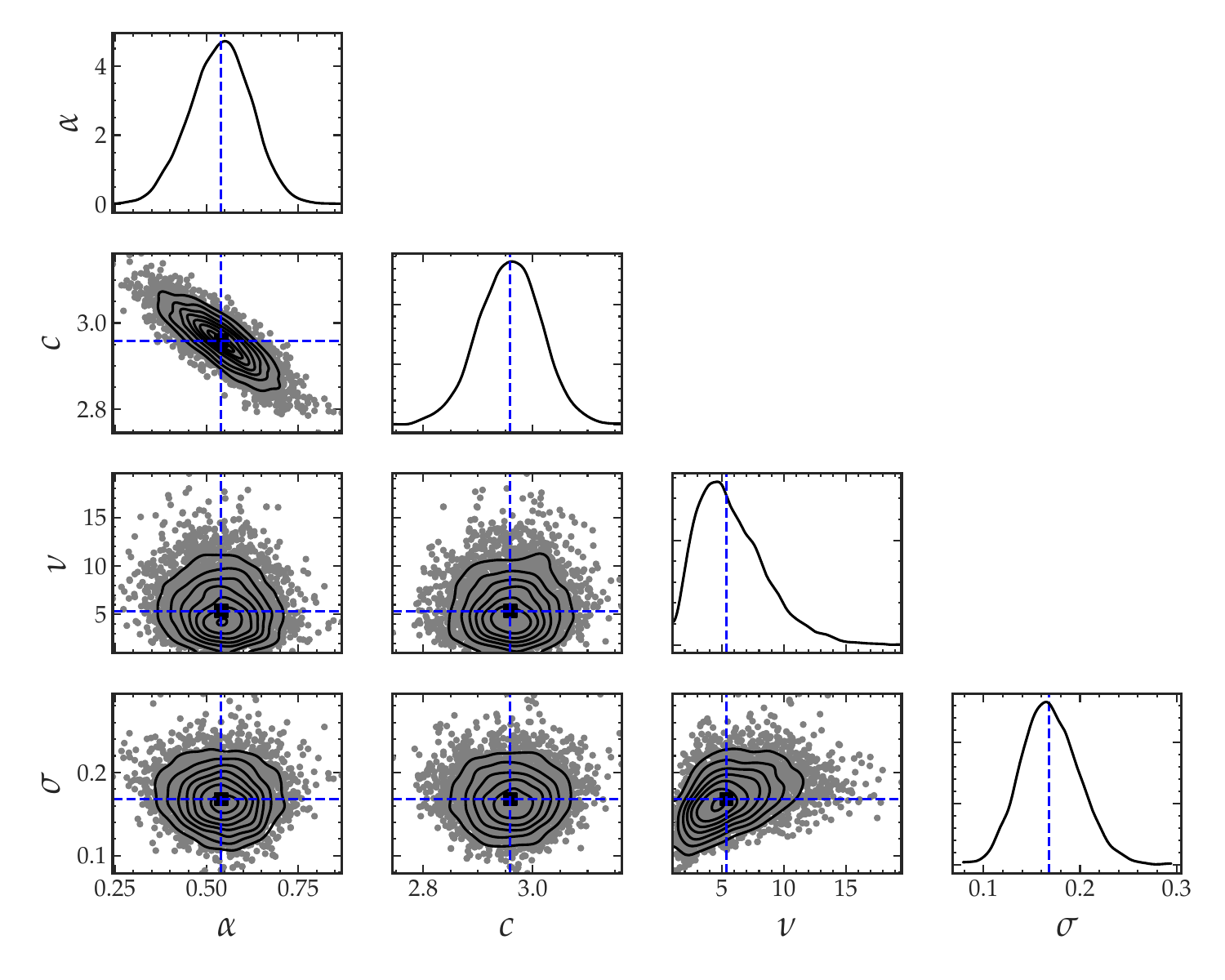}
     \end{minipage}
     \noindent
    \begin{minipage}[c]{\columnwidth}
     \centering
     \includegraphics[width=\columnwidth]{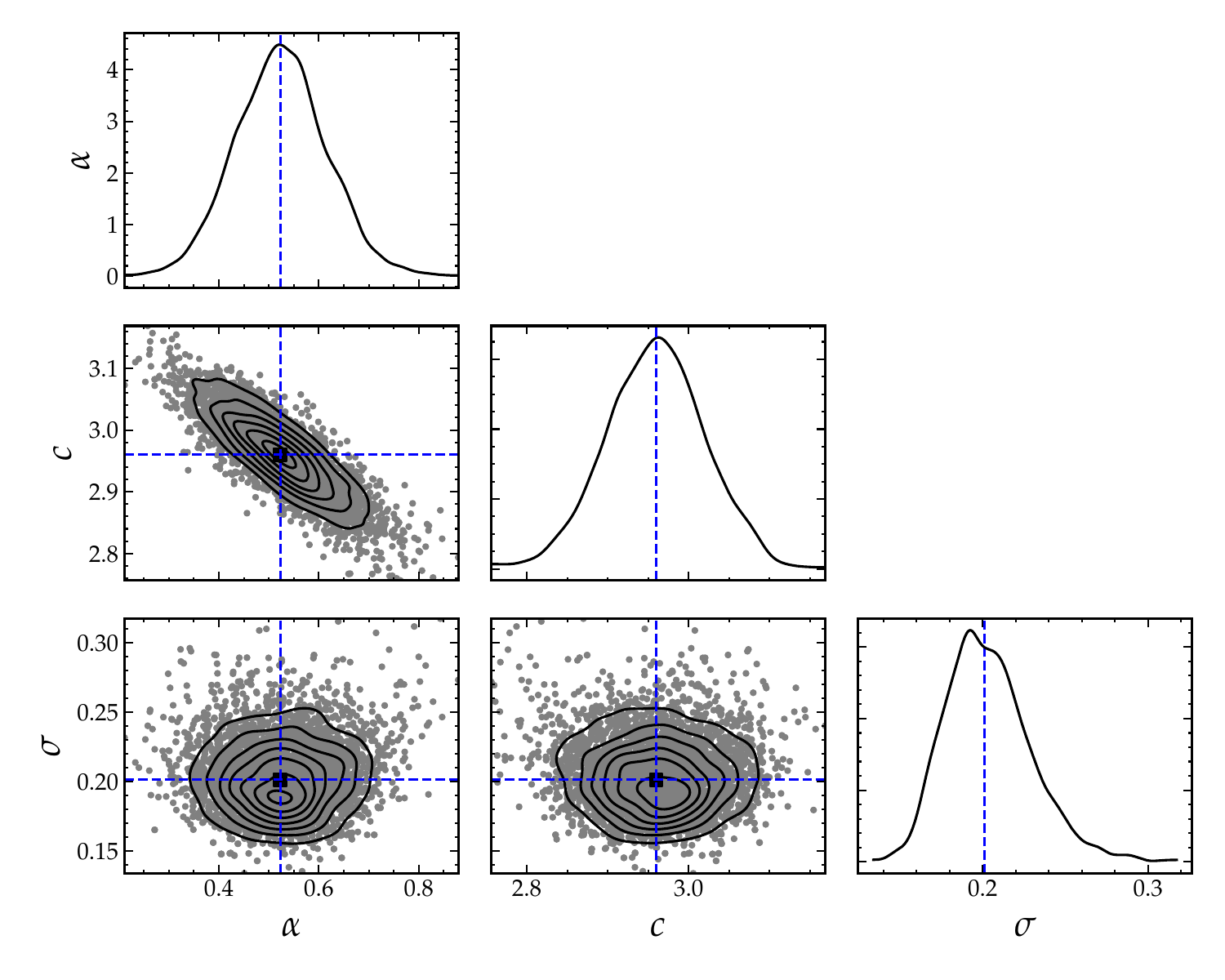}
     \end{minipage}
    \caption{The posterior distributions of the regression parameters of the \jbar$-M_{\rm bar}$ relation for the isolated $j$-sample: the posteriors in the {\it top panel} were obtained with a Student-$t$ distribution while those in the {\it bottom panel} were derived using a normal distribution.}
    \label{fig:postdist}
\end{figure}

It is worth noting that the measurement uncertainties were not accounted for in the definition of the likelihood. In principle, this consideration does not significantly impact the regression; however, it can potentially cause the {\it vertical} intrinsic scatter $\sigma$ to be overestimated. \Cref{fig:postdist} shows the posterior distributions of the $j$-sample regression, for each of the regression parameters: the slope $\alpha = \alphaj$, intercept $c = (\cj)\,\rm  dex$ and degree of freedom $\nu = \nuj$, along with the vertical intrinsic scatter $\sigma=(\sigmaj)\,\rm  dex$. All parameters exhibit unimodal distributions around their mean values, weighing in favour of the robustness of the obtained values.

For comparison, we re-performed the regression by modelling the likelihood with a normal distribution (instead of a Student-$t$ distribution): we obtained $\alpha_\mathcal{N} = 0.52 \pm 0.09$, $c_\mathcal{N} = (2.96 \pm 0.06)\,\rm dex$ and $\sigma_\mathcal{N} = (0.20 \pm 0.03)\,\rm dex$. These values are consistent with the above results, although we note that the associated intrinsic scatter is ${\sim}18\%$ larger than the previous.

\section{Convergence of angular momentum}
The analysis conducted in this paper included all galaxies from the $j$-sample, with no consideration of the convergence of their baryonic angular momentum as to not discriminate against any particular type of galaxies. In this section we distinguish between converging and non-converging galaxies following the criteria in \citet{Posti2018} and consider a galaxy converging when (i) its outermost \jbar\ values differ by less than 10\% and (ii) the slope of \jbar\ in the logarithm space is lower than half. That is, a galaxy is deemed converging when
\begin{align*}
    \frac{j_{\rm bar}(<R_N) - j_{\rm bar}(<R_{N-1})}{j_{\rm bar}(<R_N)} < 0.1 && \& && \frac{\partial \log{j_{\rm bar}(<R)}}{\partial \log{R}} < \frac{1}{2},
\end{align*}
with $R_{N-1}$ and $R_N$ the respective last two radii.

\begin{figure}
    \centering
    \includegraphics[width=\columnwidth]{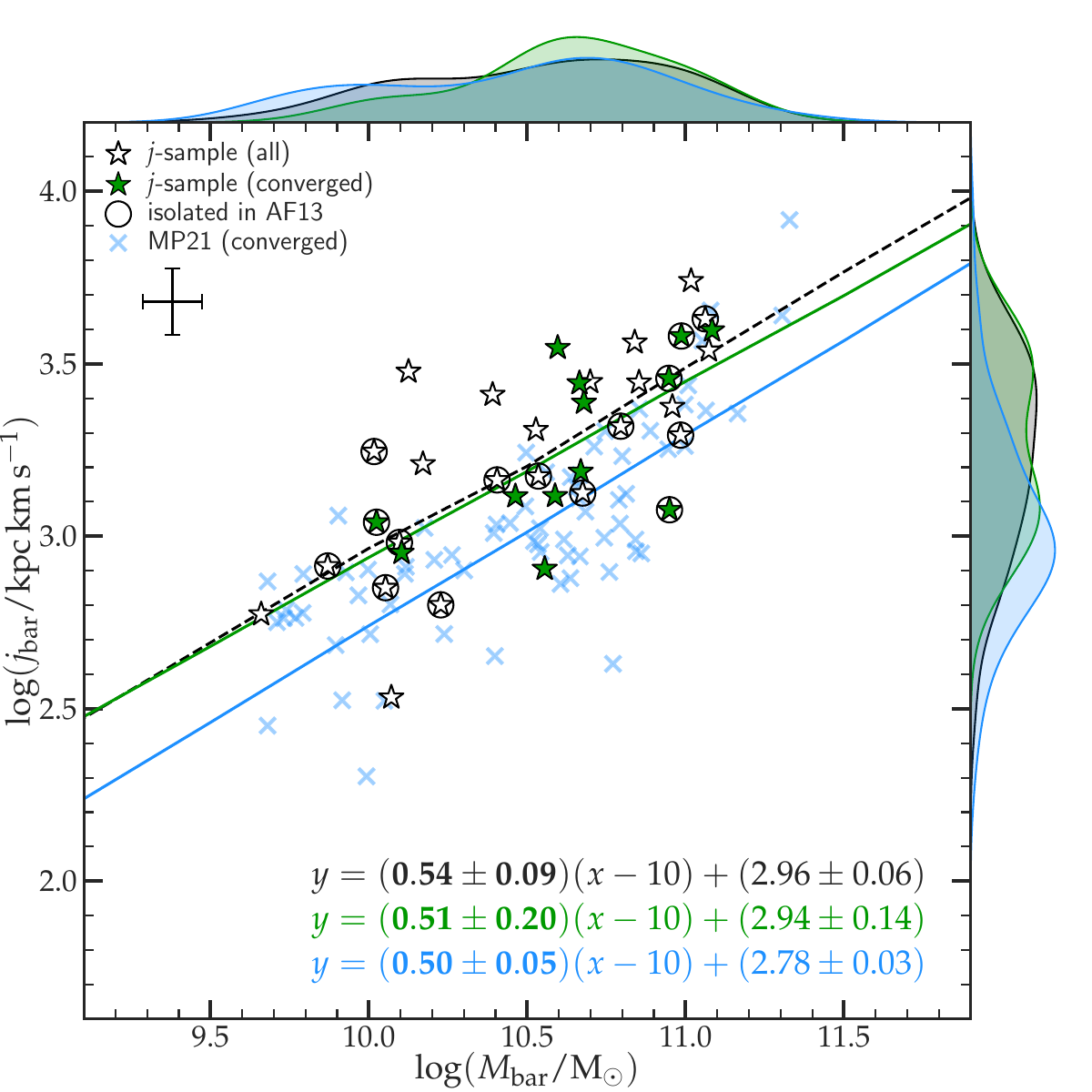}
    \vspace{-20pt}
    \caption{Same as right panel of \Cref{fig:jvsm-bar}, but distinguishing between converging and non-converging galaxies in the $j$-sample.}\label{fig:jvsm_conv}
\end{figure}

Of the 36 galaxies in the $j$-sample, only 13 fulfill the above convergence criteria. As shown in \Cref{fig:jvsm_conv}, these galaxies do not occupy a preferred position in the angular momentum space. They span the same range of baryonic masses as the non-converged galaxies and their distribution seems random. In particular, these converged galaxies do not feature among the highest \jbar\ galaxies and their line of best fit is consistent with that of the overall $j$-sample: above the converged galaxies of the \citetalias{ManceraPina2021} sample. This implies that the higher $j$ values observed in this work are independent of the convergence criteria.

\section{Table of angular momentum values}
\setlength{\tabcolsep}{2pt}
\begin{table*}
{\tiny\centering
\caption{The kinematic properties of the galaxies in the isolated $j$-sample. The columns list: (1) the CIG number; (2) the NED name; (3) the logarithm of the atomic gas mass in $\rm M_\odot$ and (4) its 1$\sigma$ error; (5) the logarithm of the stellar mass in $\rm M_\odot$ and (6) its 1$\sigma$ error; (7) the logarithm of the total baryonic mass in $\rm M_\odot$ and (8) its 1$\sigma$ error; (9) the atomic gas fraction; (10) the angular momentum of the gas in $\rm km\,s^{-1}\,kpc$ and (11) its 1$\sigma$ error; (12) the stellar angular momentum in $\rm km\,s^{-1}\,kpc$ and (13) its 1$\sigma$ error; (14) the total baryonic angular momentum in $\rm km\,s^{-1}\,kpc$ and (15) its 1$\sigma$ error; (16) the global stability parameter; (17): convergence of $j_{\rm bar}$, 1 for converged galaxies and 0 otherwise.}
\label{tb:angmom}
\begin{tabular}{l l c c c c c c c c c c c c c c c}
\hline
\hline
CIG & Other name & $\log{M_{\rm gas}}$ & $\sigma(\log{M_{\rm gas}})$ & $\log{M_{\rm star}}$ & $\sigma(\log{M_{\rm star}})$ & $\log{M_{\rm bar}}$ & $\sigma(\log{M_{\rm bar}})$ & $f_{\rm atm}$ & $\log{j_{\rm gas}}$ & $\sigma(\log{j_{\rm gas}})$ & $\log{j_{\rm star}}$ & $\sigma(\log{j_{\rm star}})$ & $\log{j_{\rm bar}}$ & $\sigma(\log{j_{\rm bar}})$ & $q$ & conv \\ 
(1) & (2) & (3) & (4) & (5) & (6) & (7) & (8) & (9) & (10) & (11) & (12) & (13) & (14) & (15) & (16) & (17)\\ 
\hline
85 & UGC 01547 & 10.07 & 0.12 & 9.20 & 0.16 & 10.12 & 0.11 & 0.88 & 3.48 & 0.06 & 3.42 & 0.07 & 3.48 & 0.06 & 0.52 & 0 \\ 
96 & NGC 0864 & 10.22 & 0.12 & 10.04 & 0.14 & 10.44 & 0.09 & 0.67 & 3.48 & 0.01 & 3.20 & 0.02 & 3.41 & 0.01 & 0.24 & 0 \\ 
102 & UGC 01886 & 10.44 & 0.12 & 10.85 & 0.12 & 10.99 & 0.09 & 0.26 & 3.95 & 0.17 & 3.63 & 0.36 & 3.74 & 0.28 & 0.12 & 0 \\ 
103 & NGC 0918 & 9.52 & 0.12 & 9.97 & 0.14 & 10.10 & 0.11 & 0.26 & 3.12 & 0.04 & 2.87 & 0.07 & 2.95 & 0.06 & 0.16 & 1 \\ 
123 & IC 0302 & 10.45 & 0.12 & 10.94 & 0.12 & 11.06 & 0.09 & 0.41 & 3.78 & 0.02 & 3.31 & 0.05 & 3.56 & 0.03 & 0.12 & 0 \\ 
134 & UGC 02883 & 9.81 & 0.12 & 10.43 & 0.12 & 10.52 & 0.10 & 0.19 & 3.55 & 0.10 & 3.25 & 0.21 & 3.32 & 0.17 & 0.15 & 0 \\ 
147 & NGC 1530 & 10.26 & 0.12 & 10.76 & 0.12 & 10.88 & 0.09 & 0.46 & 3.70 & 0.02 & 3.35 & 0.04 & 3.55 & 0.02 & 0.21 & 1 \\ 
159 & UGC 03326 & 10.21 & 0.12 & 10.68 & 0.12 & 10.81 & 0.09 & 0.18 & 3.60 & 0.04 & 3.29 & 0.08 & 3.37 & 0.07 & 0.06 & 0 \\ 
188 & UGC 03826 & 9.82 & 0.12 & 9.72 & 0.14 & 10.07 & 0.09 & 0.56 & 2.61 & 0.11 & 2.42 & 0.17 & 2.53 & 0.13 & 0.07 & 0 \\ 
232 & NGC 2532 & 10.48 & 0.12 & 10.23 & 0.12 & 10.68 & 0.09 & 0.64 & 3.19 & 0.06 & 2.97 & 0.10 & 3.12 & 0.07 & 0.06 & 0 \\ 
240 & UGC 04326 & 9.76 & 0.12 & 10.56 & 0.12 & 10.62 & 0.10 & 0.12 & 3.49 & 0.04 & 3.16 & 0.08 & 3.22 & 0.07 & 0.08 & 1 \\ 
292 & NGC 2712 & 10.02 & 0.12 & 10.14 & 0.12 & 10.38 & 0.08 & 0.41 & 3.35 & 0.02 & 2.93 & 0.05 & 3.16 & 0.03 & 0.13 & 1 \\ 
314 & NGC 2776 & 10.30 & 0.12 & 10.52 & 0.12 & 10.72 & 0.09 & 0.58 & 3.32 & 0.05 & 2.85 & 0.15 & 3.18 & 0.07 & 0.10 & 0 \\ 
329 & NGC 2862 & 10.13 & 0.12 & 10.88 & 0.12 & 10.95 & 0.10 & 0.14 & 3.79 & 0.03 & 3.11 & 0.16 & 3.30 & 0.11 & 0.05 & 0 \\ 
359 & NGC 2960 & 9.63 & 0.12 & 10.76 & 0.12 & 10.79 & 0.11 & 0.05 & 3.64 & 0.02 & 3.47 & 0.04 & 3.48 & 0.03 & 0.08 & 0 \\ 
361 & NGC 2955 & 10.61 & 0.12 & 10.94 & 0.12 & 11.11 & 0.09 & 0.34 & 3.81 & 0.06 & 3.42 & 0.14 & 3.60 & 0.10 & 0.08 & 1 \\ 
421 & UGC 05700 & 10.39 & 0.12 & 10.66 & 0.12 & 10.85 & 0.09 & 0.34 & 3.62 & 0.20 & 3.34 & 0.38 & 3.46 & 0.29 & 0.09 & 0 \\ 
463 & UGC 06162 & 9.89 & 0.12 & 9.70 & 0.14 & 10.11 & 0.09 & 0.62 & 3.08 & 0.11 & 2.75 & 0.24 & 2.98 & 0.14 & 0.18 & 0 \\ 
512 & UGC 06903 & 9.75 & 0.12 & 9.75 & 0.14 & 10.05 & 0.09 & 0.50 & 2.93 & 0.03 & 2.75 & 0.05 & 2.85 & 0.04 & 0.15 & 0 \\ 
551 & UGC 07941 & 9.87 & 0.12 & 9.24 & 0.16 & 9.96 & 0.10 & 0.70 & 3.12 & 0.04 & 2.74 & 0.09 & 3.04 & 0.05 & 0.24 & 1 \\ 
553 & NGC 4719 & 10.26 & 0.12 & 10.83 & 0.12 & 10.93 & 0.10 & 0.20 & 3.37 & 0.07 & 2.96 & 0.17 & 3.08 & 0.13 & 0.03 & 1 \\ 
571 & NGC 4964 & 9.38 & 0.04 & 9.70 & 0.14 & 9.87 & 0.10 & 0.32 & 3.14 & 0.07 & 2.78 & 0.15 & 2.93 & 0.11 & 0.27 & 0 \\ 
581 & NGC 5081 & 10.49 & 0.12 & 10.88 & 0.12 & 11.03 & 0.09 & 0.27 & 3.93 & 0.02 & 3.44 & 0.07 & 3.63 & 0.05 & 0.09 & 0 \\ 
604 & NGC 5377 & 9.59 & 0.12 & 10.60 & 0.12 & 10.64 & 0.11 & 0.10 & 3.47 & 0.06 & 3.05 & 0.15 & 3.12 & 0.13 & 0.08 & 1 \\ 
616 & UGC 09088 & 10.25 & 0.12 & 10.88 & 0.12 & 10.97 & 0.10 & 0.18 & 3.88 & 0.04 & 3.49 & 0.09 & 3.60 & 0.07 & 0.09 & 1 \\ 
626 & NGC 5584 & 9.96 & 0.12 & 10.05 & 0.14 & 10.31 & 0.10 & 0.54 & 2.87 & 0.08 & 2.71 & 0.12 & 2.80 & 0.10 & 0.09 & 0 \\ 
660 & UGC 09730 & 9.51 & 0.12 & 9.14 & 0.16 & 9.66 & 0.10 & 0.71 & 2.84 & 0.07 & 2.55 & 0.13 & 2.77 & 0.08 & 0.30 & 0 \\ 
676 & UGC 09853 & 10.16 & 0.12 & 10.62 & 0.12 & 10.75 & 0.09 & 0.23 & 3.55 & 0.05 & 3.21 & 0.11 & 3.31 & 0.09 & 0.08 & 0 \\ 
736 & NGC 6118 & 9.91 & 0.12 & 10.38 & 0.12 & 10.51 & 0.09 & 0.28 & 3.28 & 0.02 & 3.03 & 0.04 & 3.11 & 0.03 & 0.10 & 1 \\ 
744 & UGC 10437 & 9.92 & 0.12 & 9.32 & 0.16 & 10.02 & 0.10 & 0.80 & 3.27 & 0.05 & 3.13 & 0.07 & 3.25 & 0.06 & 0.39 & 0 \\ 
983 & UGC 12173 & 10.05 & 0.12 & 10.58 & 0.12 & 10.69 & 0.09 & 0.23 & 3.62 & 0.08 & 3.28 & 0.17 & 3.39 & 0.13 & 0.12 & 1 \\ 
988 & UGC 12190 & 10.48 & 0.12 & 10.83 & 0.12 & 10.99 & 0.09 & 0.25 & 3.78 & 0.05 & 3.42 & 0.12 & 3.54 & 0.09 & 0.07 & 0 \\ 
1000 & UGC 12260 & 9.90 & 0.12 & 9.83 & 0.14 & 10.17 & 0.09 & 0.54 & 3.33 & 0.09 & 2.97 & 0.21 & 3.20 & 0.13 & 0.25 & 0 \\ 
1004 & NGC 7479 & 10.22 & 0.12 & 10.52 & 0.12 & 10.70 & 0.09 & 0.36 & 3.61 & 0.03 & 3.31 & 0.07 & 3.44 & 0.05 & 0.14 & 1 \\ 
1006 & UGC 12372 & 9.96 & 0.12 & 10.42 & 0.12 & 10.55 & 0.09 & 0.25 & 3.05 & 0.20 & 2.85 & 0.31 & 2.91 & 0.27 & 0.05 & 1 \\ 
1019 & NGC 7664 & 10.34 & 0.12 & 10.34 & 0.12 & 10.64 & 0.08 & 0.44 & 3.62 & 0.07 & 3.26 & 0.16 & 3.46 & 0.10 & 0.13 & 0 \\ 
\hline
\end{tabular}}
\end{table*}

\section{Moment maps}\label{app:velfields}
\begin{figure*}
    \includegraphics[height=\textheight]{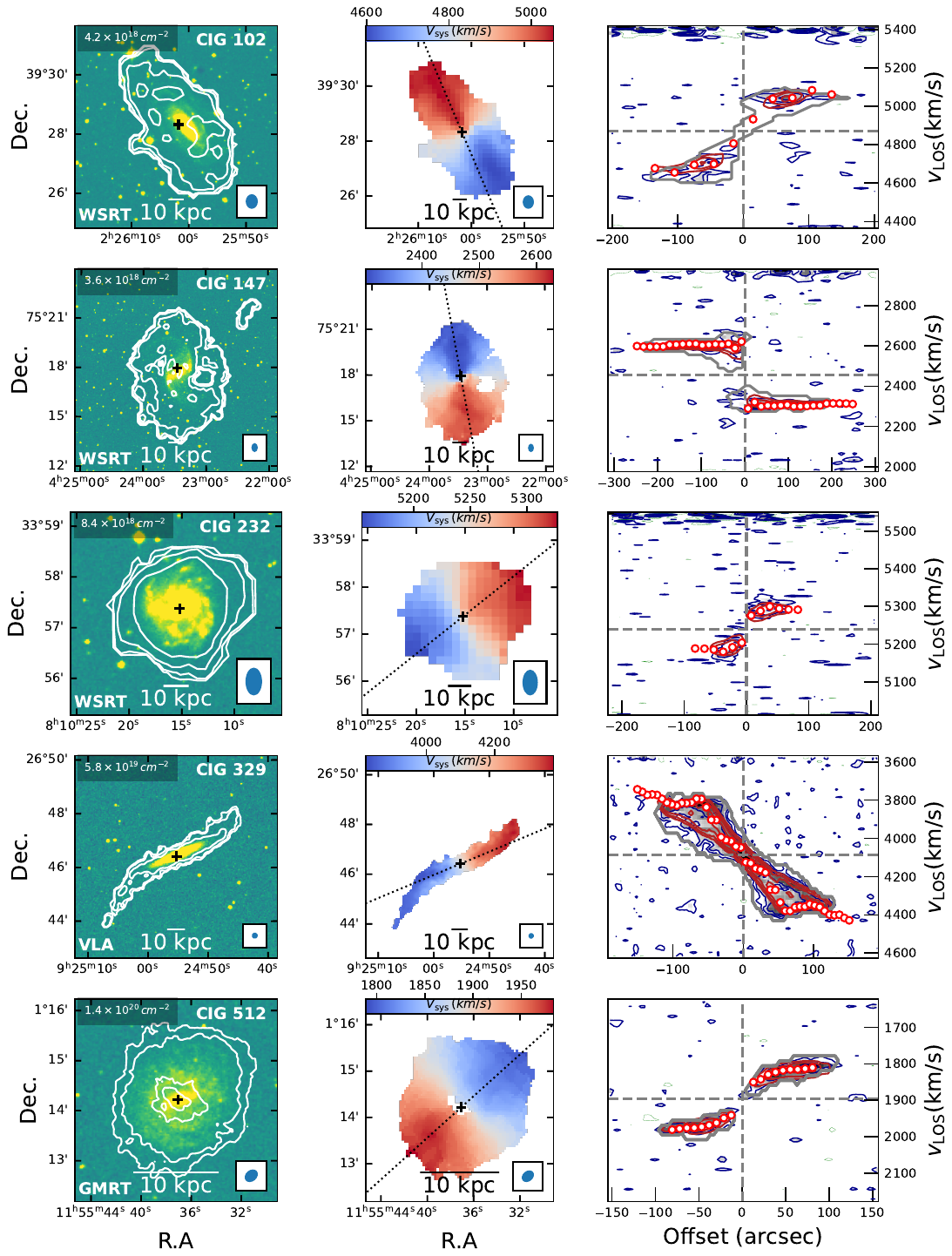}
    \vspace{-50pt}
    \caption[]{Integrated \hi\ maps ({\it left}), velocity fields ({\it centre}) and position-velocity diagrams ({\it right}) of AMIGA galaxies. The telescope used for each galaxy is given in the bottom left corner of the left panel.}
    \label{fig:velmap}
\end{figure*}

Each row of \Cref{fig:velmap} contains the moment maps and position-velocity diagram of a CIG galaxy of the $j$-sample. The {\it left panel} shows the integrated \hi\ maps as contours overlaid on DSS2 {\it r}-band images. The CIG ID is given in the top right corner, the lowest column density contour level (taken at $3\sigma$) in the top left corner, the telescope whose data was used in the bottom left corner and a representation of the beam in the bottom right corner. The scale is also shown in the bottom center of the panel. The contours increment as $3\sigma\times2^{n}$ with $n=0,2,4,\dots$. The {\it middle panel} shows the velocity fields obtained from first moments, with the velocity values given by the horizontal bar above the panel. Finally, the rotation curve ({\it red circles}) is overlaid on the position-velocity diagram in the {\it right panel}. The blue contours represent the data, the red contours the model and the thick gray contours the mask within which the model was computed. The figure only includes five selected galaxies, the full sample is shown in the online supplementary material.

\section{Rotation curves}\label{app:rotcur}
\begin{figure*}
    \centering
    \includegraphics[height=\textheight]{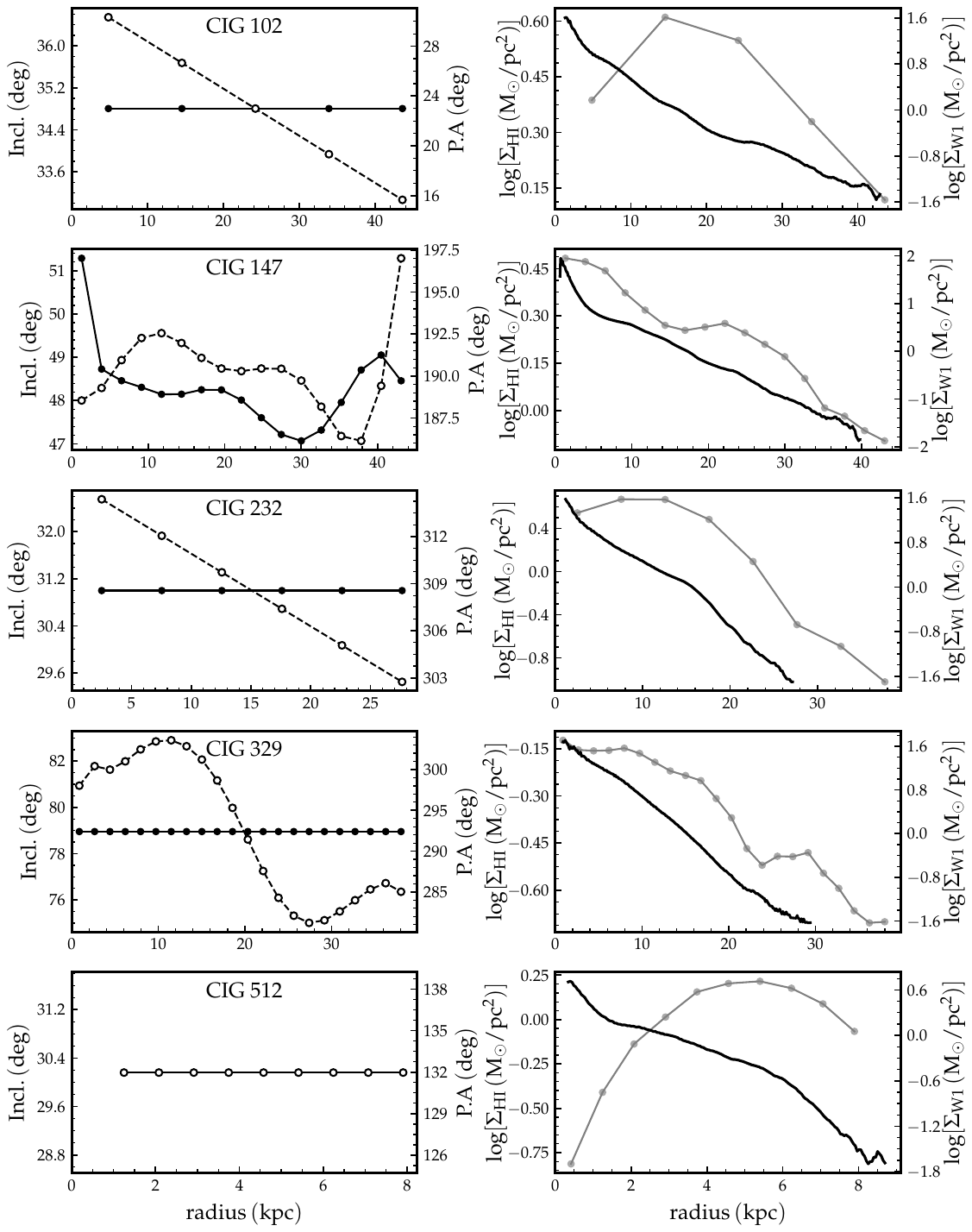}
    \vspace{-50pt}
    \caption{{\it Left:} variation of the inclination ({\it solid line}) and position angle ({\it dashed line}) at different radii of the galaxy. {\it Right:} the resulting \hi\ surface density profile {\it gray curve} as well as the \wise\ W1 surface density profile.}\label{fig:rotpar}
\end{figure*}

\Cref{fig:rotpar} shows the variations of the orientation parameters (the inclination and position angle) and the optical and \hi\ surface density profiles for the five galaxies included in \Cref{fig:velmap}. The full sample is given in the online supplementary material.

In \Cref{fig:rotcur} we show the rotation curves of all galaxies in the $j$-sample. The horizontal dashed line denotes the average velocity $V_{\rm flat}$ along the flat part of the rotation curve. $V_{\rm flat}$ is estimated following the method prescribed in \citet{Lelli2016a}; that is, starting at the outermost radius $N$ of the rotation curve, we evaluate the mean velocity 
\begin{equation}
    \bar{V} = \frac{1}{2}(V_N + V_{N-1}).
\end{equation}
As long as the velocity of the next point $N-2$ is such that
\begin{equation}
    \frac{V_{N-2} - \bar{V}}{\bar{V}} \leq \varepsilon
\end{equation}
(where $\varepsilon = 0.1$ is the maximum variation allowed in the flat part), the iteration continues to the next point and so on. When the above condition breaks, we take $V_{\rm flat} = \bar{V}$, and estimate its error as
\begin{equation}
    \delta_{V_{\rm flat}} = \sqrt{\frac{1}{N}\sum_{n}^{N}\delta_{V_n}^2 + \left(\frac{V_{\rm flat}}{\tan{(\rm incl.)}}\delta_{\rm incl.}\right)^2 + \delta^2_{\bar{V}}},
\end{equation}
function of the inclination {\it incl.}

\begin{figure*}
    \includegraphics[height=\textheight]{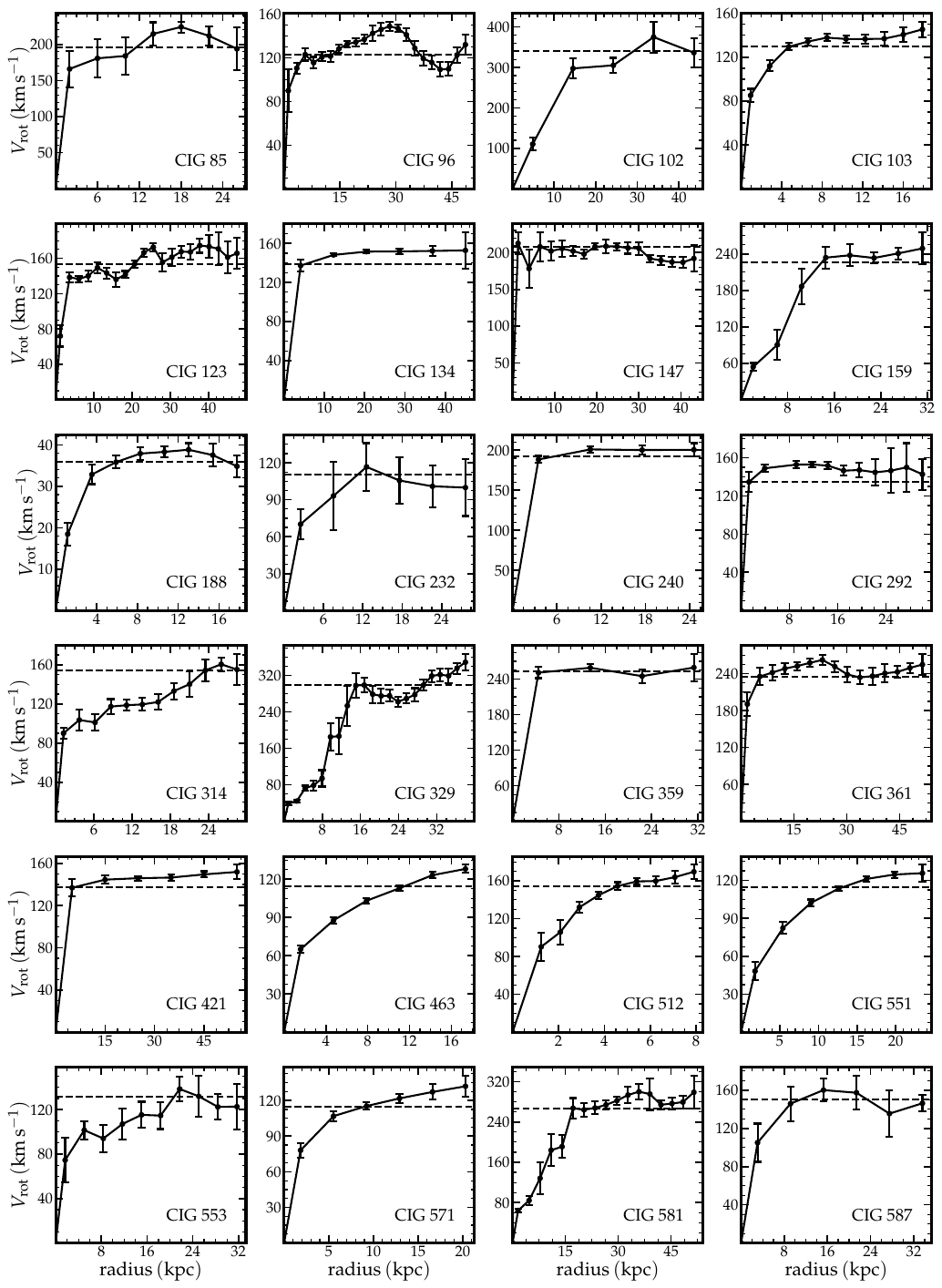}
    \vspace{-50pt}
    \caption[]{Rotation curves of the $j$-sample galaxies. The dashed horizontal lines represent the $V_{\rm flat}$, the average flat velocity.}
    \label{fig:rotcur}
\end{figure*}

\begin{figure*}
    \ContinuedFloat
    \includegraphics[height=\textheight]{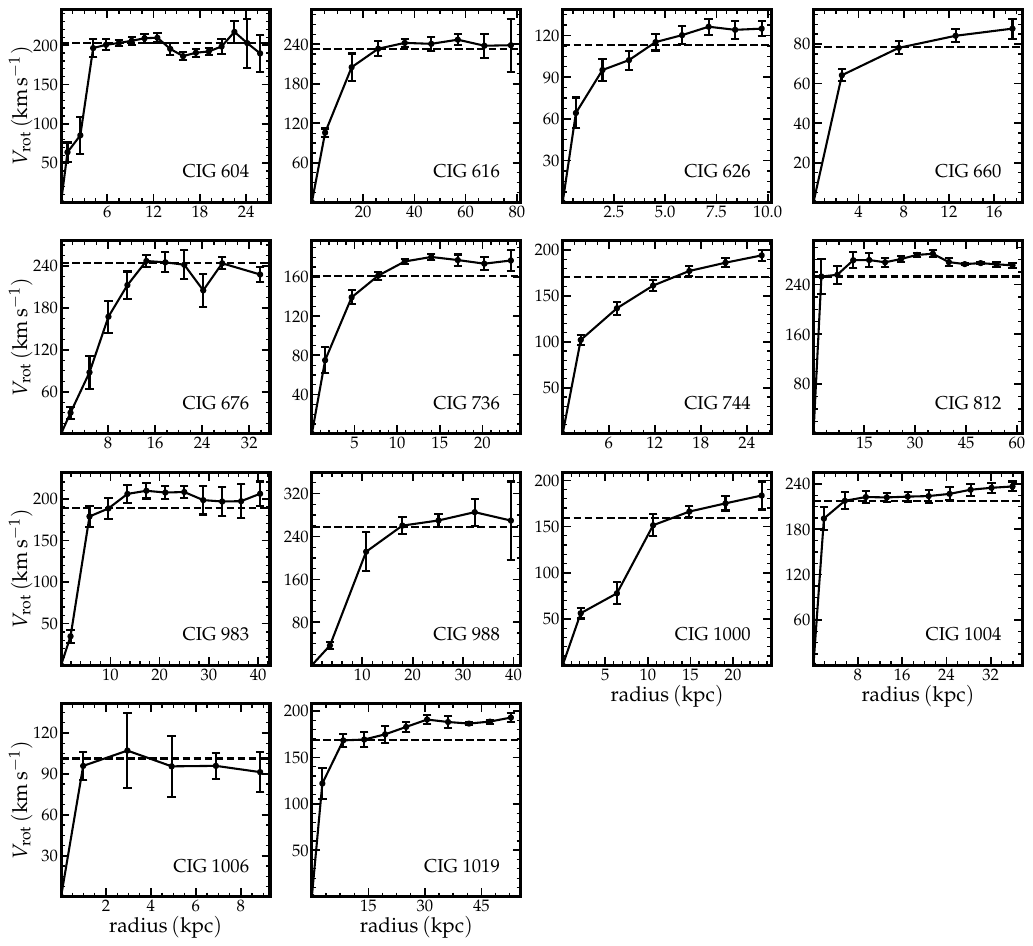}
    \vspace{-50pt}
    \caption{\emph{continued from previous page}}
\end{figure*}

\bsp	
\label{lastpage}
\end{document}